\documentclass[twocolumn]{aastex62}
\usepackage{amsfonts}
\usepackage{amsmath}
\usepackage{mathrsfs}
\usepackage{epsfig}

\def\kpc{\,\mathrm{kpc}}
\def\km{\,\mathrm{km}}
\def\m{\,\mathrm{m}}

\def\GeV{\,\mathrm{GeV}}

\def\GV{\,\mathrm{GV}}
\def\TV{\,\mathrm{TV}}

\def\cm{\,\mathrm{cm}}
\def\s{\,\mathrm{s}}
\def\p{\,\mathrm{p}}
\def\pbar{\,\bar{\text{p}}}
\def\He{\,\mathrm{He}}
\def\C{\,\mathrm{C}}

\def\O{\,\mathrm{O}}
\def\Li{\,\mathrm{Li}}
\def\Be{\,\mathrm{Be}}
\def\B{\,\mathrm{B}}
\def\A{\,\mathrm{A}}
\def\sr{\,\mathrm{sr}}
\def\pbar{\,\bar{\text{p}}}
\def\pbarp{\,\bar{\text{p}}/\text{p}}

\def\phip{\phi_{\mathrm{p}}}
\def\phipbar{\phi_{\pbar}}
\def\phihe{\phi_\mathrm{He}}
\def\phic{\phi_\mathrm{C}}
\def\phin{\phi_\mathrm{N}}
\def\phio{\phi_\mathrm{O}}
\def\phili{\phi_\mathrm{Li}}
\def\phibe{\phi_\mathrm{Be}}
\def\phib{\phi_\mathrm{B}}

\def\cpbar{c^{\mathrm{sec}}_{\mathrm{\bar{\text{p}}}}}
\def\che{c^{\mathrm{pri}}_{\mathrm{He}}}
\def\cc{c^{\mathrm{pri}}_{\mathrm{C}}}
\def\cnp{c^{\mathrm{pri}}_{\mathrm{N}}}
\def\co{c^{\mathrm{pri}}_{\mathrm{O}}}
\def\cli{c^{\mathrm{sec}}_{\mathrm{Li}}}
\def\cbe{c^{\mathrm{sec}}_{\mathrm{Be}}}
\def\cb{c^{\mathrm{sec}}_{\mathrm{B}}}
\def\cns{c^{\mathrm{sec}}_{\mathrm{N}}}

\def\csec{c_{i}^{\mathrm{sec}}}

\def\Rbr{\,R_{\mathrm{br}}}

\def\R{\mathcal{R}}

\def\phipbar{\phi_{\pbar}}

\received{}
\revised{}
\accepted{}
\submitjournal{ApJ}

\shorttitle{Spectra hardening}
\shortauthors{Niu et al.}

\begin{document}

\title{Bayesian analysis of the hardening in AMS-02 nuclei spectra}

\correspondingauthor{Jia-Shu Niu}
\email{jsniu@sxu.edu.cn, jsniu@itp.ac.cn}

\author[0000-0001-5232-9500]{Jia-Shu Niu}
\affil{Institute of Theoretical Physics, Shanxi University, Taiyuan, 030006, China}
\affil{CAS Key Laboratory of Theoretical Physics, Institute of Theoretical Physics, Chinese Academy of Sciences, Beijing, 100190, China}

\author{Tianjun Li}
\affiliation{CAS Key Laboratory of Theoretical Physics, Institute of Theoretical Physics, Chinese Academy of Sciences, Beijing, 100190, China}
\affiliation{School of Physical Sciences, University of Chinese Academy of Sciences, No.~19A Yuquan Road, Beijing 100049, China}

\author[0000-0001-6027-4562]{Hui-Fang Xue}
\affiliation{Department of Astronomy, Beijing Normal University, Beijing 100875, China}



\begin{abstract}

  Based on the precise nuclei data released by AMS-02, we study the spectra hardening of both the primary (proton, helium, carbon, oxygen, and the primary component of nitrogen) and the secondary (anti-proton, lithium, beryllium, boron and the secondary component of nitrogen) cosmic ray (CR) nuclei. With the diffusion-reacceleration model, we consider two schemes to reproduce the hardening in the spectra: (i) A high-rigidity break in primary source injection; (ii) A high-rigidity break in diffusion coefficient. The global fitting results show that both schemes could reproduce the spectra hardening in current status. More precise multi-TV data (especially the data of secondary CR species) is needed if one wants to distinguish these two schemes. In our global fitting, each of the nuclei species is allocated an independent solar modulation potential and a re-scale factor (which accounts for the isotopic abundance for primary nuclei species and uncertainties of production cross section or inhomogeneity of CR sources and propagation for secondary nuclei species). The fitting values of these two parameter classes show us some hints on some new directions in CR physics. All the fitted re-scale factors of primary nuclei species have values that systematically smaller than 1.0, while that of secondary nuclei species are systematically larger than 1.0. Moreover, both the re-scale factor and solar modulation potential of beryllium have values which are obviously different from other species. This might indicate that beryllium has the specificity not only on its propagation in the heliosphere, but also on its production cross section. All these new results should be seriously studied in the future.

\end{abstract}



\section{Introduction}

Understanding the spectral features in cosmic rays (CRs) is of fundamental importance for studying their origin and propagation. Great progress in cosmic ray (CR) spectrum measurement has been made in recent years with a new generation of space-borne and ground-based experiments in operation. The fine structure of spectral hardening for primary nuclei at $\sim 300 \GV$ was observed by ATIC-2 \citep{ATIC2006}, CREAM \citep{CREAM2010}, PAMELA \citep{PAMELA2011}, and AMS-02 \citep{AMS02_proton,AMS02_helium}. 

Recently, AMS-02 has released the energy spectra of He, C, and O \citep{AMS02_He_C_O}, which confirmed the spectral hardening of CR primary nuclei. Moreover, the subsequently released energy spectra of Li, Be, and B \citep{AMS02_Li_Be_B} show that the secondary nuclei spectra harden even more than that of the primary ones at a few hundred GV.  After that, the released nitrogen spectrum \citep{AMS02_N} (which is made up of both primary and secondary components) shows that the spectral index rapidly hardens at high rigidities and become identical to the spectral indices of primary He, C, and O CRs above $\sim$ 700 GV.  Because the secondary CR particles are produced in collisions of primary CR particles with interstellar medium (ISM), combining these data together would provide us an excellent opportunity to study the hardening of the CR nuclei spectra quantitatively. 

Some previous works have proposed different solutions to this problem: (i) adding a new break in high-energy region ($\sim 300 \GV$) to the injection spectra (see, e.g., \citet{Korsmeier2016,Boschini2017,Niu2017_dampe1,Niu2017_dampe2,Zhu2018}); (ii) adding a new in high-rigidity break to the diffusion coefficient (see, e.g., \citet{Genolini2017}); (iii) inhomogeneous diffusion (see, e.g., \citet{Blasi2012,Tomassetti2012,Tomassetti2015apjl01,Tomassetti2015prd,Feng2016,Guo2018}); (iv) the superposition of local and distant sources (see, e.g., \citet{Vladimirov2012,Bernard2013,Thoudam2013,Tomassetti2015apjl02,Kachelriess2015,Kawanaka2018}).

In this work, we perform a global fitting on these primary and secondary nuclei spectra from AMS-02. Two schemes are considered: (i) the hardening of the observed spectra comes from the sources (the breaks are already present in the spectra after the CR particles accelerated at the sources) --  a new high-rigidity break is added in the primary source injections (Scheme I); (ii) the hardening of the observed spectra comes from the propagation --  a new high-rigidity break is added in the diffusion coefficient (Scheme II). We hope that the precise spectra data from AMS-02 would give us a clear result, at least a tendency.

The paper is organized as follows. We first list the setups in Section \ref{sec:setups}. The fitting results are give in Section \ref{sec:fitting_results}. Then we  give a test of the best-fit results in Section \ref{sec:test} and present some discussions and conclusions in Section \ref{sec:dis_con}.

\section{Setups}
\label{sec:setups}

In this section, we list some of the most important setups in this work, more detailed similar configurations could be found in \citet{Niu2018},  and some important differences are listed and discussed in sub-Section \ref{subsec:diff}.

\subsection{Model}

As the setup in our previous work \citep{Niu2018,Niu2017_dampe1}, we use independent primary source spectra settings for proton and other nuclei species because of the significant difference observed in the slopes of proton and other nuclei species when $Z > 1$ \citep{AMS02_proton,AMS02_helium,AMS02_He_C_O}.  Moreover, in our calculation, a cylindrically symmetric geometry is assumed to describe the CR propagation in the galaxy, with a fixed maximum radius $r = 20 \kpc$.

\subsubsection{Propagation Model}

We consider the diffusion-reacceleration model in the global fitting,  which is widely used and consistent with the  AMS-02 nuclei data (see, e.g., \citet{Niu2018,Yuan2017,Yuan2018}).  $v_A$ is used to characterize the reacceleration, and  $z_h$ represents the half-height of the propagation region in the galaxy for the cylindrical coordinate system.  In the whole propagation region,  a uniform diffusion coefficient is used which depends on CR particles' rigidity.

In Scheme I, the diffusion coefficient is parametrized as 
\begin{equation}
\label{eq:diffusion_coefficient_1}
D_{xx}(R) = D_0\beta \left( \frac{R}{R_{0}} \right)^{\delta}~,~\,
\end{equation}
where $\beta$ is the velocity of the particle in unit of light speed $c$,  $R_{0}$ is the reference rigidity (4 GV),
and $R\equiv pc/Ze$ is the rigidity.

For Scheme II, the diffusion coefficient is parametrized as 
\begin{equation}
\label{eq:diffusion_coefficient_2}
D_{xx}(R)=  D_{0} \cdot \beta \left(\frac{\Rbr}{R_{0}}\right) \times \left\{
  \begin{array}{ll}
    \left( \dfrac{R}{\Rbr} \right)^{\delta_{1}} & R \le \Rbr\\
    \left( \dfrac{R}{\Rbr} \right)^{\delta_{2}} & R > \Rbr
  \end{array}
  \right.,
\end{equation}
where $\Rbr$ is the high-rigidity break, $\delta_{1}$ and $\delta_{2}$ are the diffusion slopes below and above the break.

\subsubsection{Primary Sources}

The primary source injection spectra of all kinds of nuclei are assumed to be a broken power law form. In Scheme I, it is represented as:
\begin{equation}
  q_{\mathrm{i}}=  N_{\mathrm{i}} \times \left\{ \begin{array}{ll}
\left( \dfrac{R}{R\mathrm{_{A1}}} \right)^{-\nu_{\A1}} & R \leq R_{\A1}\\
\left( \dfrac{R}{R\mathrm{_{A1}}} \right)^{-\nu_{\A2}} & R_{\A1} < R \leq R_{\A2} \\
\left( \dfrac{R}{R\mathrm{_{A2}}} \right)^{-\nu_{\A3}} \left( \dfrac{R\mathrm{_{A2}}}{R\mathrm{_{A1}}} \right)^{-\nu_{\A2}} & R > R_{\A2} 
  \end{array}
  \right.,
  \label{eq:injection_spectra_1}
\end{equation}
where $\mathrm{i}$ denotes the species of nuclei, $N_\mathrm{i}$ is the normalization constant proportional to the relative abundance of the corresponding nuclei, and $\nu_{\A}=\nu_{\A1}(\nu_{\A2}, \nu_{\A3})$ for the nucleus rigidity $R$ in the region divided by 2 breaks at the reference rigidity $R_{\A1}$ and $R_{\A2}$. In this work, we use independent proton injection spectrum, and the corresponding parameters are $R_{\p1}$, $R_{\p2}$, $\nu_{\p1}$, $\nu_{\p2}$,  and $\nu_{\p3}$. All the $Z > 1$ nuclei are assumed to have the same value of injection parameters.

For Scheme II, we have
\begin{equation}
  q_{\mathrm{i}} =  N_{\mathrm{i}} \times \left\{ \begin{array}{ll}
\left( \dfrac{R}{R\mathrm{_{A}}} \right)^{-\nu_{\A1}}  & R \leq R_{\A}\\
\left( \dfrac{R}{R\mathrm{_{A}}} \right)^{-\nu_{\A2}} & R > R_{\A} 
  \end{array}
  \right.,
  \label{eq:injection_spectra_1}
\end{equation}
which are described by one break at the rigidity $R_{\A}$ ($R_{\p}$) and two slopes below ($\nu_{\A1}$ or $\nu_{\p1}$) and above ($\nu_{\A2}$ or $\nu_{\p2}$) it.

\subsubsection{Solar modulation}

We adopt the force-field approximation \citep{Gleeson1968} to describe the effects of solar modulation in the solar system, which contains only one parameter the so-called solar-modulation potential $\phi$. Considering the charge-sign and suspected nuclei species dependence of the solar modulation which is represented in  previous fitting \citep{Niu2018}, we adopt $\phip$, $\phihe$, $\phic$, $\phio$, $\phipbar$, $\phili$, $\phibe$, and $\phib$ to modulate the proton, He, C, O, $\pbar$, Li, Be, and B nuclei data, respectively. This would give us the limitation of force-field approximation as a simple and effective theory to describe the solar modulation on local interstellar spectra (LIS).  Moreover, it would show us the differences of the propagation between different nuclei species in the heliosphere.

\subsubsection{Numerical tools}

The public code  {\sc galprop} v56 \footnote{http://galprop.stanford.edu} \citep{Strong1998,Moskalenko2002,Strong2001,Moskalenko2003,Ptuskin2006} is used to solve the diffusion equation numerically.
 In {\sc galprop}, the primary source (injection) isotopic abundances are determined by fitting to the data from ACE at $\sim$ 200 MeV/nucleon assuming a propagation model \citep{Wiedenbeck2001,Wiedenbeck2008}. This configuration appears some discrepancies when fit to some new data covering high energy regions \citep{Johannesson2016}. As a result, we use factors $\che$, $\cc$, and $\co$ to re-scale the helium-4 (with a default abundance of $7.199 \times 10^{4}$), carbon-12 (with a default abundance of $2.819 \times 10^{3}$), and oxygen-16 (with a default abundance of $3.822 \times 10^{3}$) abundances.\footnote{In {\sc galprop}, the abundance of proton is fixed to $10^{6}$. All the other primary nuclei abundances are set to be a value which is relative to the proton abundance.} At the same time, $\cpbar$, $\cli$, $\cbe$, and $\cb$ are employed to re-scale the secondary CR nuclei species ($\pbar$, Li, Be, and B). On the one hand, these values could partially account for the production cross section uncertainties of these species (like that for $\pbar$ in \citet{Niu2018}). On the other hand, these values could also partially account for the local inhomogeneity of the CR sources and propagation. 
Here, we expect that a constant factor is a simple  assumption, which would  help us to get a better fitting result.

\subsection{Data Sets and Parameters}

In this work, the proton flux (from AMS-02 and CREAM \citep{AMS02_proton,CREAM2010}), helium flux (from AMS-02 and CREAM \citep{AMS02_He_C_O,CREAM2010}), carbon flux (from AMS-02 \citep{AMS02_He_C_O}), oxygen flux (from AMS-02 \citep{AMS02_He_C_O}), anti-proton flux (from AMS-02 \citep{AMS02_pbar_proton}), lithium flux (from AMS-02 \citep{AMS02_Li_Be_B}), beryllium flux (from AMS-02 \citep{AMS02_Li_Be_B}), and boron flux (from AMS-02 \citep{AMS02_Li_Be_B}) are added in the global fitting data set. The CREAM data is  used as the supplement of the AMS-02 data because it is more compatible with the AMS-02 data when $R \gtrsim 1 \TV$. The errors used in our global fitting are the quadratic sum of statistical and systematic errors.

Altogether, the data set in our global fitting is 
 \begin{align*}
   D = &\{D^{\text{AMS-02}}_{\p}, D^{\text{AMS-02}}_{\He},  D^{\text{AMS-02}}_{\C}, D^{\text{AMS-02}}_{\O}, \\
   &D^{\text{AMS-02}}_{\pbar}, D^{\text{AMS-02}}_{\Li}, D^{\text{AMS-02}}_{\Be},  D^{\text{AMS-02}}_{\B}, \\
&D^{\text{CREAM}}_{\p}, D^{\text{CREAM}}_{\He}  \}~.
 \end{align*}

The parameter sets for Scheme I is 
\begin{align*}
  \boldsymbol{\theta}_{1} =  &\{ D_{0}, \delta, z_{h}, v_{A}, | \\
                             & R_{\p1}, R_{\p2}, \nu_{\p1}, \nu_{\p2}, \nu_{\p3}, \\
  & R_{\A1},  R_{\A2}, \nu_{\A1}, \nu_{\A2}, \nu_{\A3}, | \\
  & N_{\p},  \che, \cc, \co, \cpbar, \cli, \cbe, \cb, | \\
& \phip, \phihe, \phic, \phio, \phipbar, \phili, \phibe, \phib \}~,
\end{align*}
for Scheme II is 
\begin{align*}
  \boldsymbol{\theta}_{2} =  &\{ D_{0}, \Rbr,  \delta_{1}, \delta_{2}, z_{h}, v_{A}, | \\
  & R_{\p}, \nu_{\p1}, \nu_{\p2},  R_{\A},   \nu_{\A1}, \nu_{\A2}, | \\
  & N_{\p}, \che, \cc, \co,   \cpbar, \cli, \cbe, \cb, | \\
& \phip, \phihe, \phic, \phio, \phipbar, \phili, \phibe, \phib \}~.
\end{align*}

These parameters can be separated into four classes: the propagation parameters, the primary source injection parameters, the normalization parameters\footnote{This class includes the normalization parameters of proton ($N_{\p}$), and the $c_{i}$s (which are called re-scale factors hereinafter).}, and the solar modulation potentials. Their priors are chosen to be uniform distributions  with the prior intervals given in Tables \ref{tab:scheme_params_I} and \ref{tab:scheme_params_II}.

\subsection{Comparing with previous work}
\label{subsec:diff}

Compared with our previous work \citep{Niu2018}, we list some of the most important updates as follows:

\begin{itemize}
\item[(i)] Because we have proved that the diffusion-reacceleration model (DR) is good enough to reproduce current AMS-02 nuclei spectra (proton, helium, B/C, and $\pbarp$), we use it as the unique propagation model in this work, and do not consider other models.

\item[(ii)] In previous work, the rigidity of the data we used in our global fitting: for the primary CR nuclei spectra (proton and helium), $\lesssim 3 \TV$; for the ratio of secondary to primary nuclei species (B/C and $\pbarp$), $\lesssim 1 \TV$. Although we got a trend in the results which shows the hardening in the spectra of the primary CR nuclei, there are not enough data points in high energy regions to do a quantitatively study on the spectra hardening. In this work, based on the primary CR nuclei spectra reaching up to 200 TV and the secondary CR nuclei spectra reaching up to 3 TV with multiple species, it is possible to study the hardening in these spectra efficiently. With the new data set, we add a new high-rigidity break in diffusion coefficient and primary source injection to account for the hardening of the spectra.
  
\item[(iii)] In previous work, we have shown that it is difficult to describe the solar modulation effects on different CR species by using a single solar modulation potential. Consequently, we assign an independent solar modulation potential to each of the CR species in this work.
  
\item[(iv)] Benefited from  the data set we used in this work (spectra other than ratios), we employ an independent re-scale factor to each of the CR species in this work. This provide us an opportunity to study the abundances of the primary sources and inhomogeneity of CR sources and propagation or production cross section of secondary CR species.
  
\item[(v)] In this work, we use an updated version of {\sc galprop} (v56), which have added some new features and updated some of the cross section data in it (see more details in \citet{galprop2017}).
  
\item[(vi)] In previous work, we focus on the comparison between different propagation models and the constraints on some special parameters ($D_{0}$ and $z_{h}$) by different data sets. In this work, we focus on the origin of the hardening in the spectra based on one data set.
\end{itemize}

All the differences discussed above are summarized in Table \ref{tab:diff_works}.

\begin{table*}[!htbp]
\caption{The main differences between our previous work \citep{Niu2018} and this work. }

\begin{center}
\begin{tabular}{l||l|l}
  \hline\hline\hline
  &Previous work &This work \\
  \hline\hline
  Propagation model &DR\footnote{Diffusion-reacceleration model} and DRC\footnote{Diffusion-reacceleration-convection model}   &DR  \\
  \hline
  Diffusion coefficient &Power law without breaks   &One break power law  \\
  \hline
  Primary source injection &One break power law  &Two breaks power law  \\
  \hline
  Re-scale factor &$\che$, $\cpbar$    &$\che, \cc, \co, \cpbar, \cli, \cbe, \cb$  \\
  \hline
  Solar modulation potential &A single $\phi$   &$\phip, \phihe, \phic, \phio, \phipbar, \phili, \phibe, \phib$  \\
  \hline
  &  &$\{D^{\text{AMS-02}}_{\p}, D^{\text{AMS-02}}_{\He},  D^{\text{AMS-02}}_{\C}, D^{\text{AMS-02}}_{\O},$  \\
  Data set &$\{D^{\text{AMS-02}}_{\p}, D^{\text{AMS-02}}_{\He},  D^{\text{AMS-02}}_{\pbarp}, D^{\text{AMS-02}}_{\mathrm{B/C}} \}$   &$D^{\text{AMS-02}}_{\pbar}, D^{\text{AMS-02}}_{\Li}, D^{\text{AMS-02}}_{\Be},  D^{\text{AMS-02}}_{\B},$  \\
  &  &$D^{\text{CREAM}}_{\p}, D^{\text{CREAM}}_{\He} \}$  \\
\hline\hline\hline
\end{tabular}
\end{center}
\label{tab:diff_works}
\end{table*}

\section{Fitting Results}
\label{sec:fitting_results}

\begin{figure*}[!htbp]
  \centering
  \includegraphics[width=0.43\textwidth]{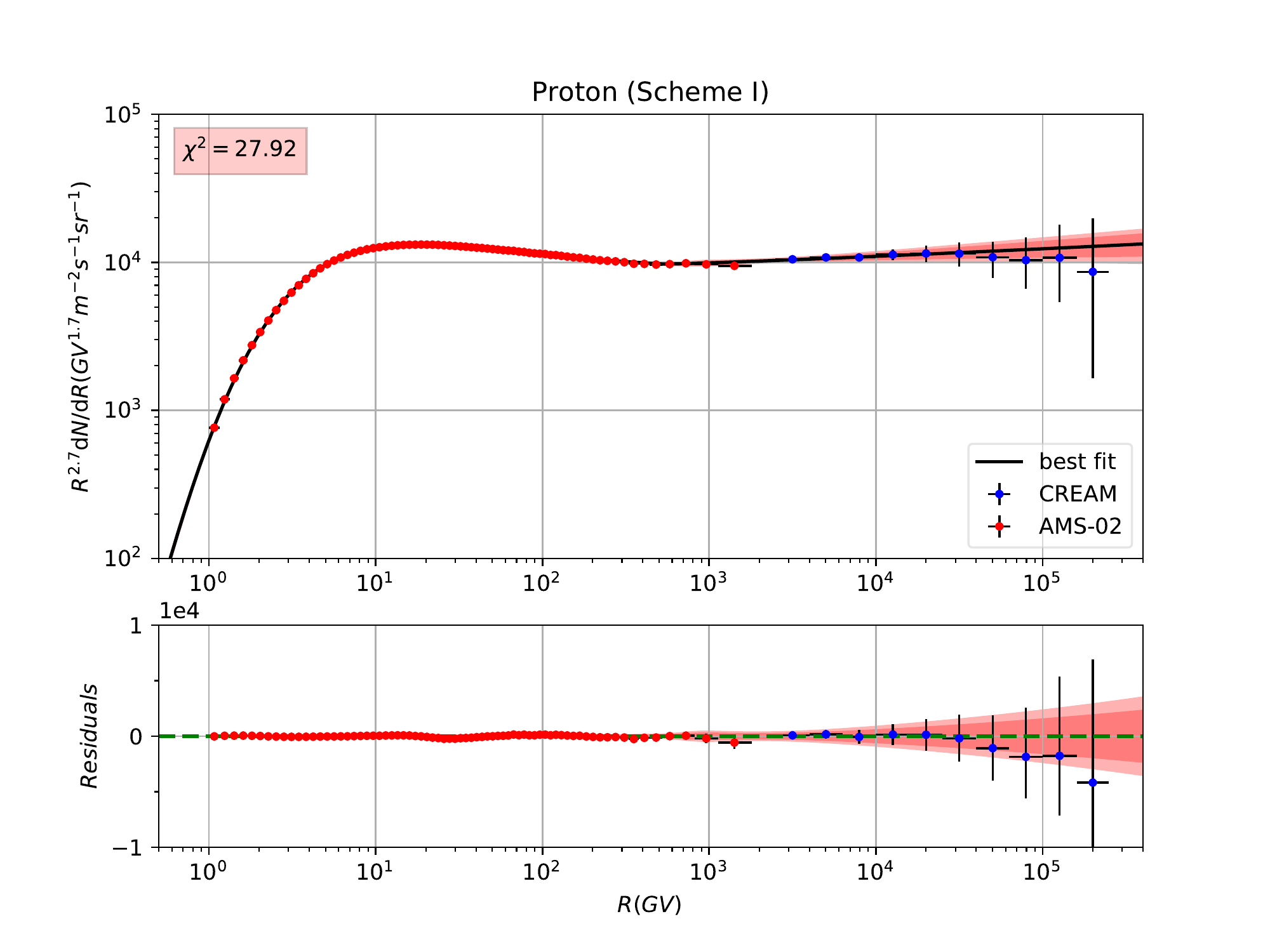}
  \includegraphics[width=0.43\textwidth]{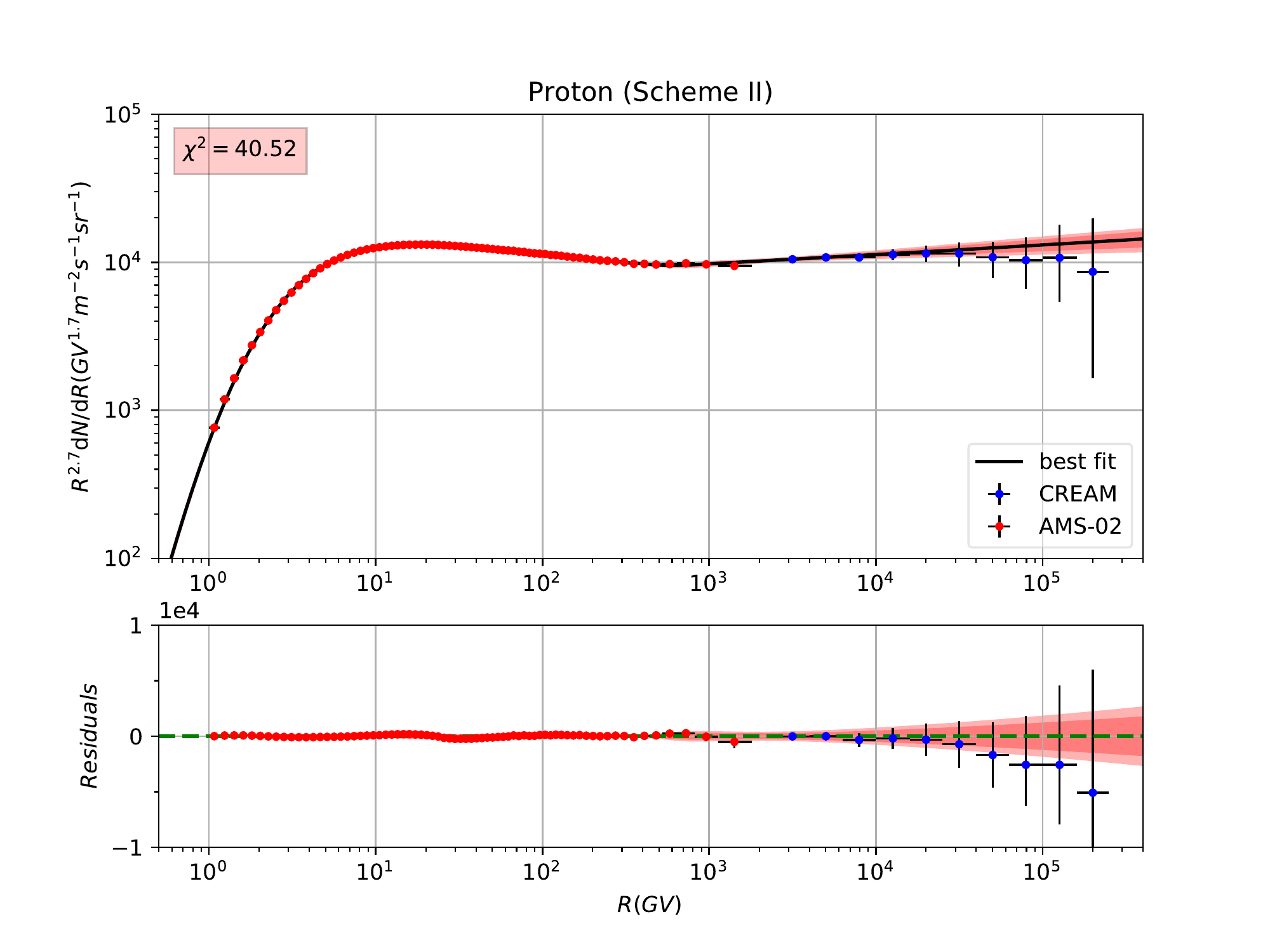}
  \includegraphics[width=0.43\textwidth]{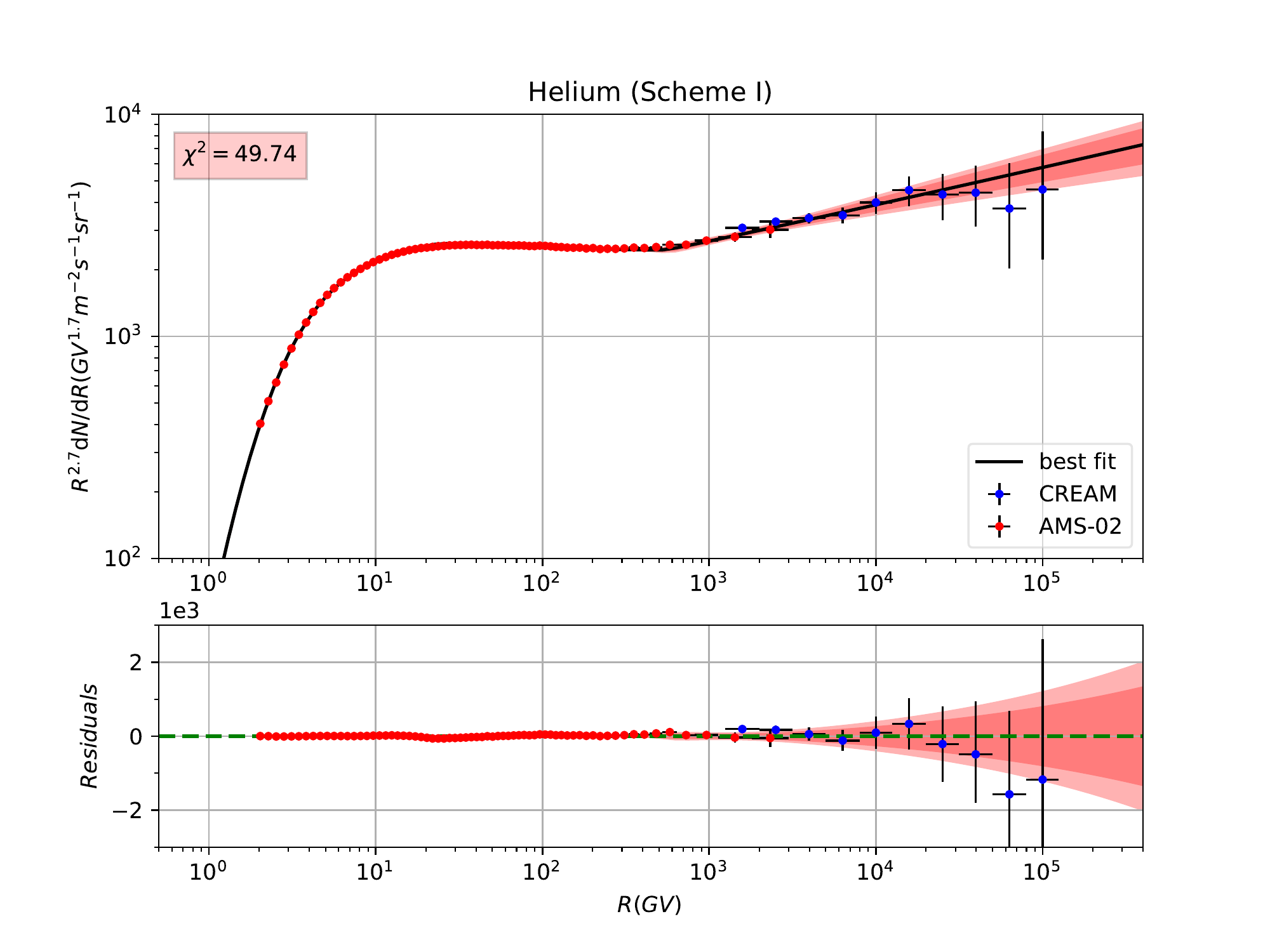}
  \includegraphics[width=0.43\textwidth]{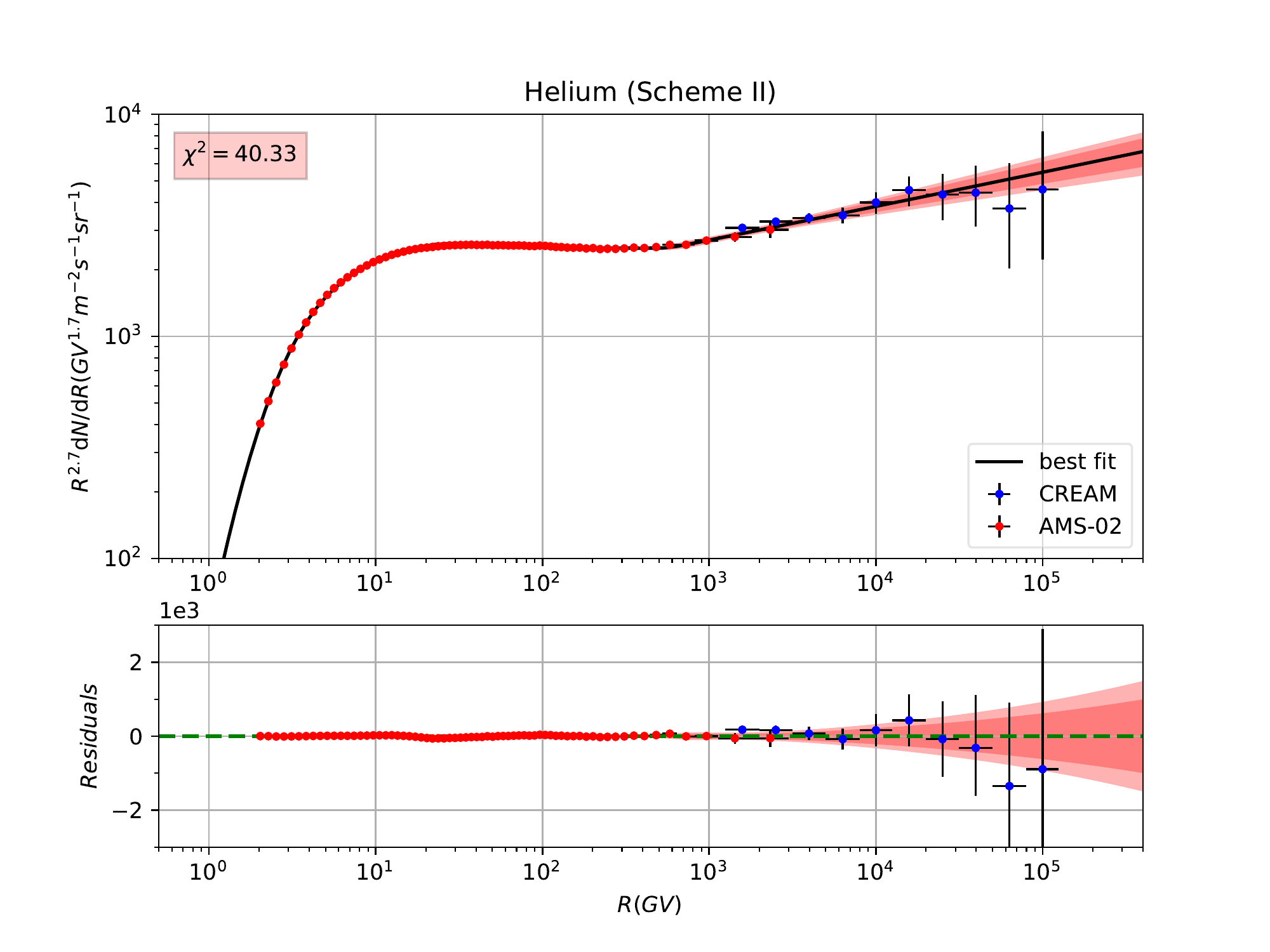}
  \includegraphics[width=0.43\textwidth]{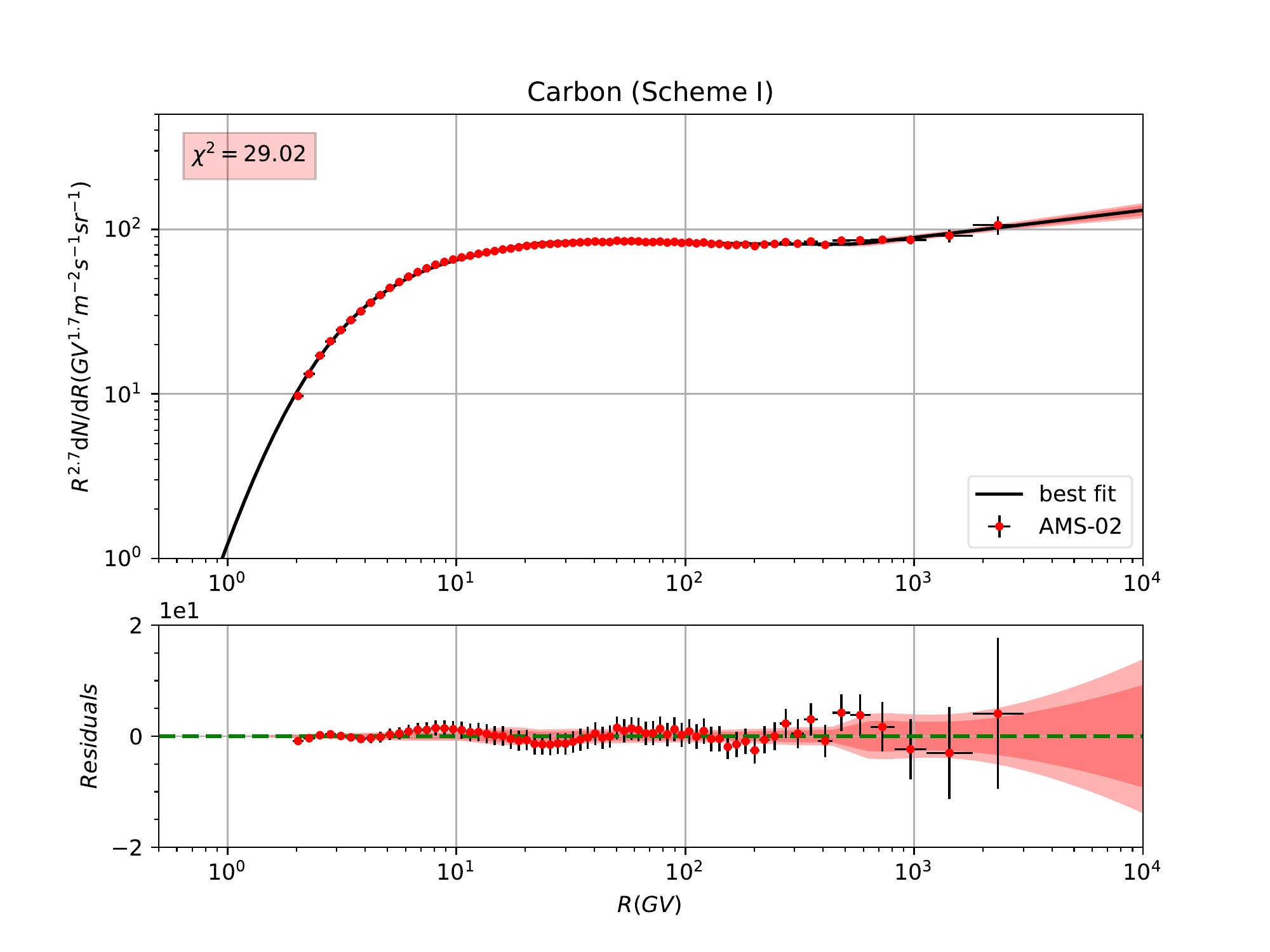}
  \includegraphics[width=0.43\textwidth]{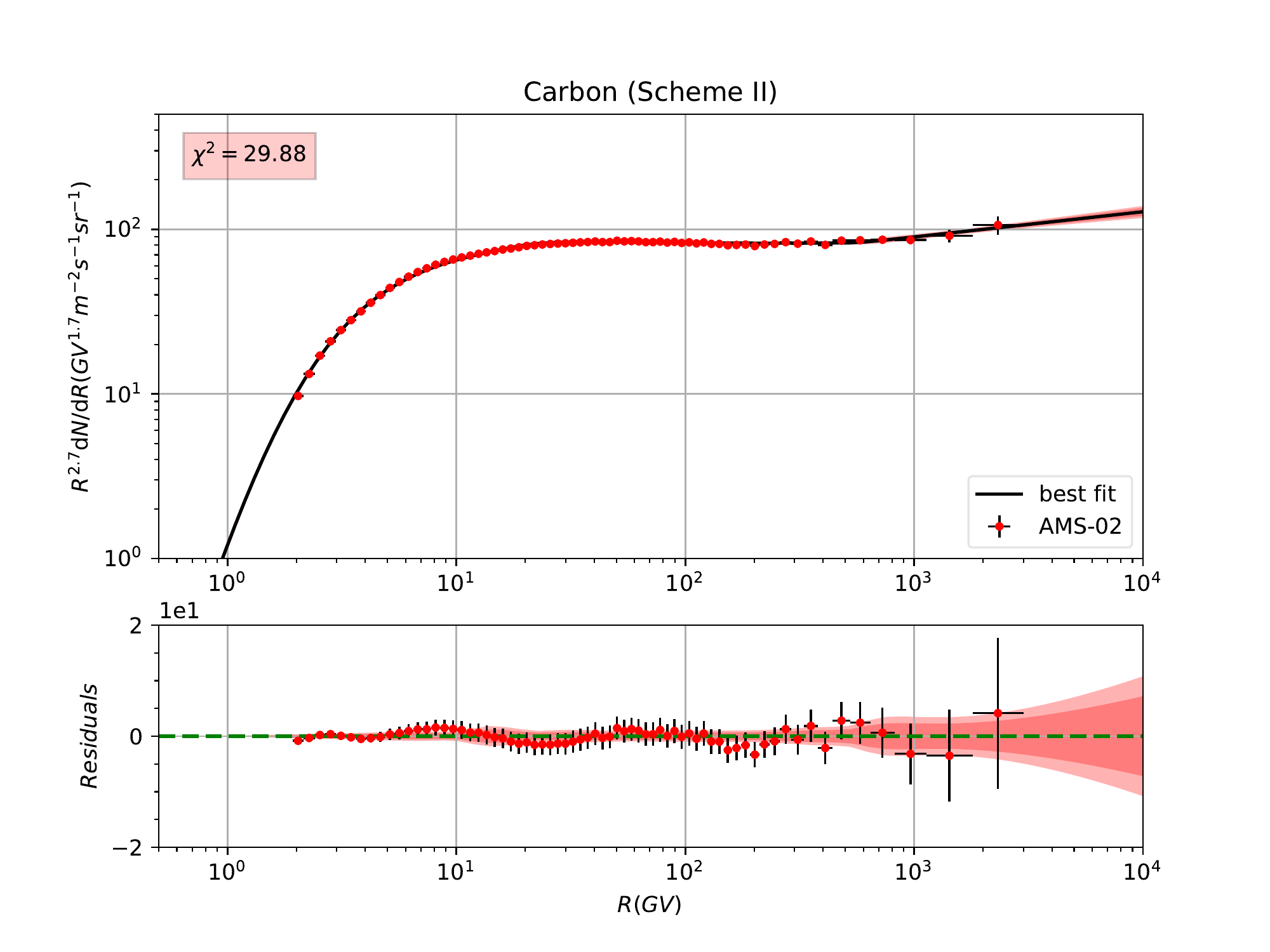}
  \includegraphics[width=0.43\textwidth]{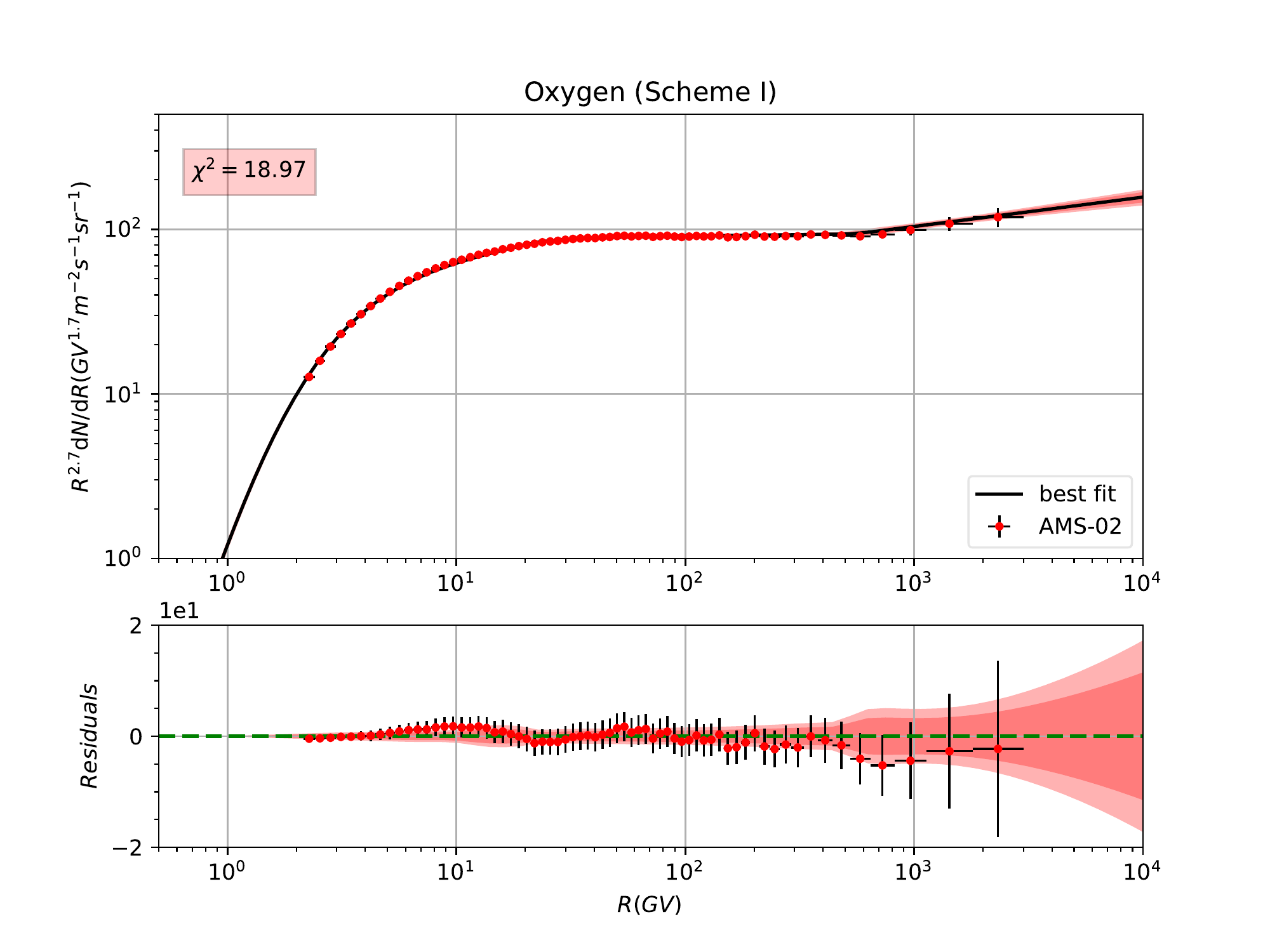}
  \includegraphics[width=0.43\textwidth]{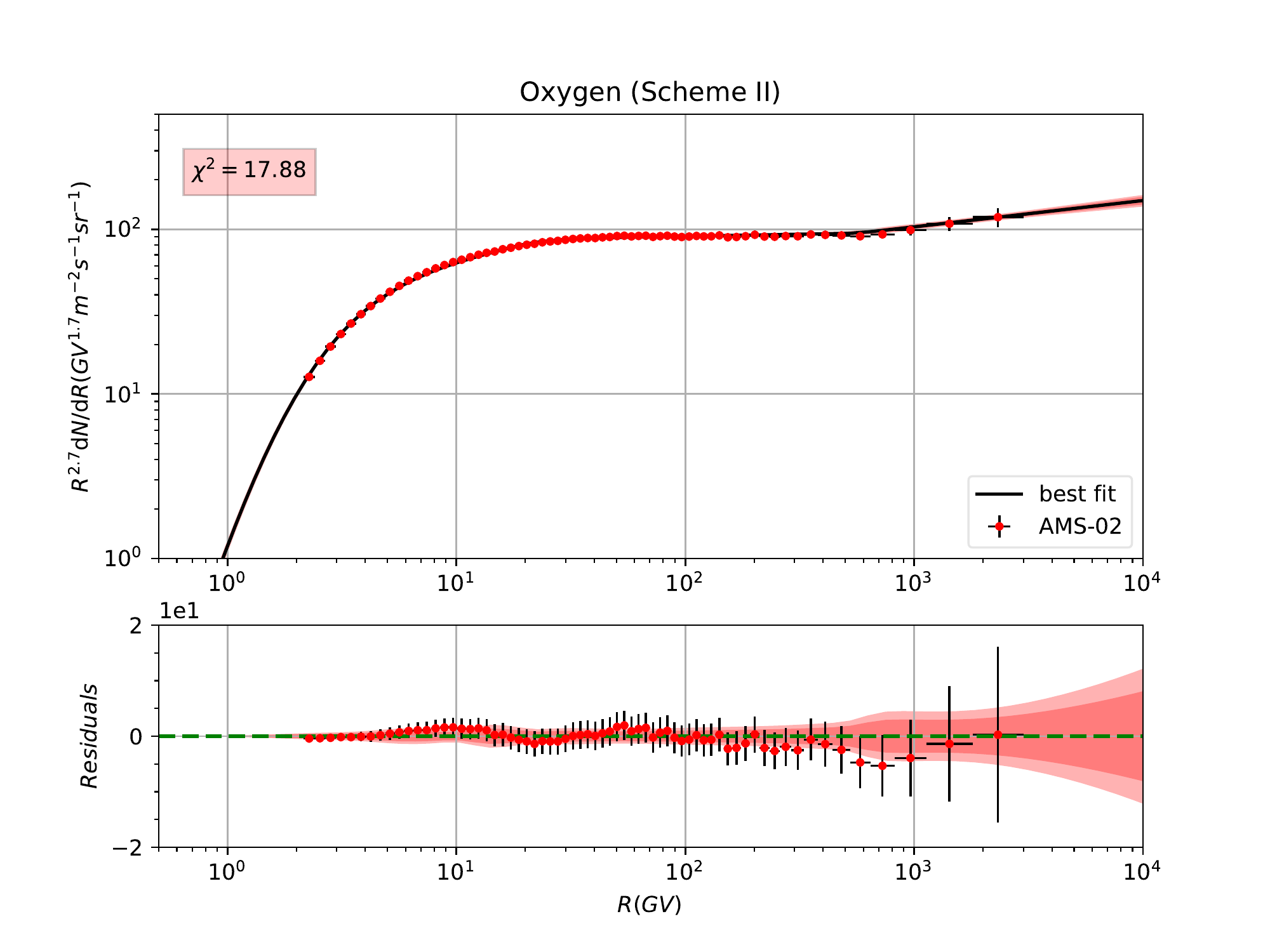}
  \caption{The global fitting results and the corresponding residuals to the primary nuclei flux (proton flux, helium flux, carbon flux and oxygen flux) for two schemes. The $2\sigma$ (deep red) and $3\sigma$ (light red) bounds are also shown in the figures. The relevant $\chi^{2}$ of each nuclei species is given in the sub-figures as well.}
\label{fig:primary_results}
\end{figure*}

\begin{figure*}[!htbp]
  \centering
  \includegraphics[width=0.43\textwidth]{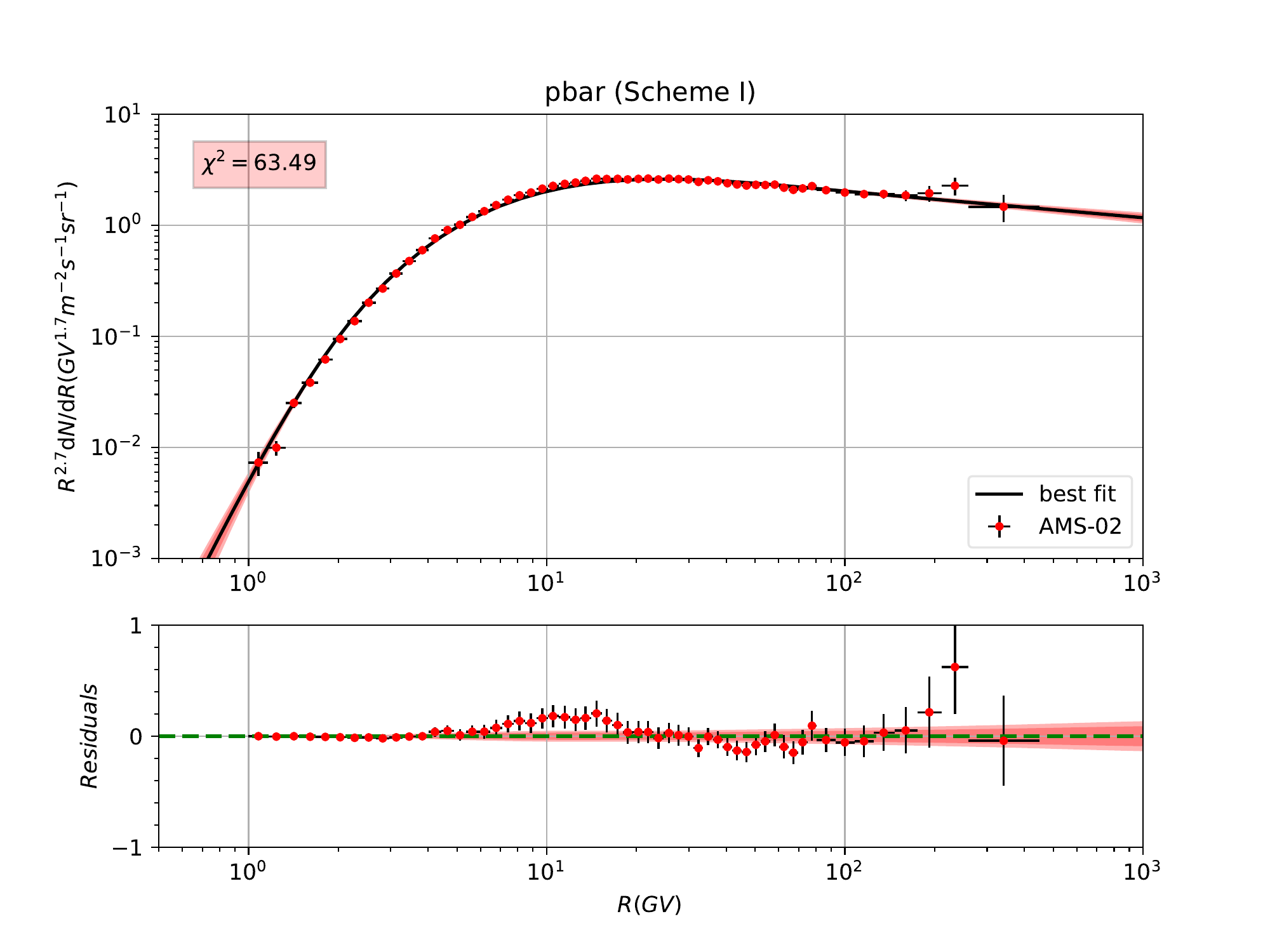}
  \includegraphics[width=0.43\textwidth]{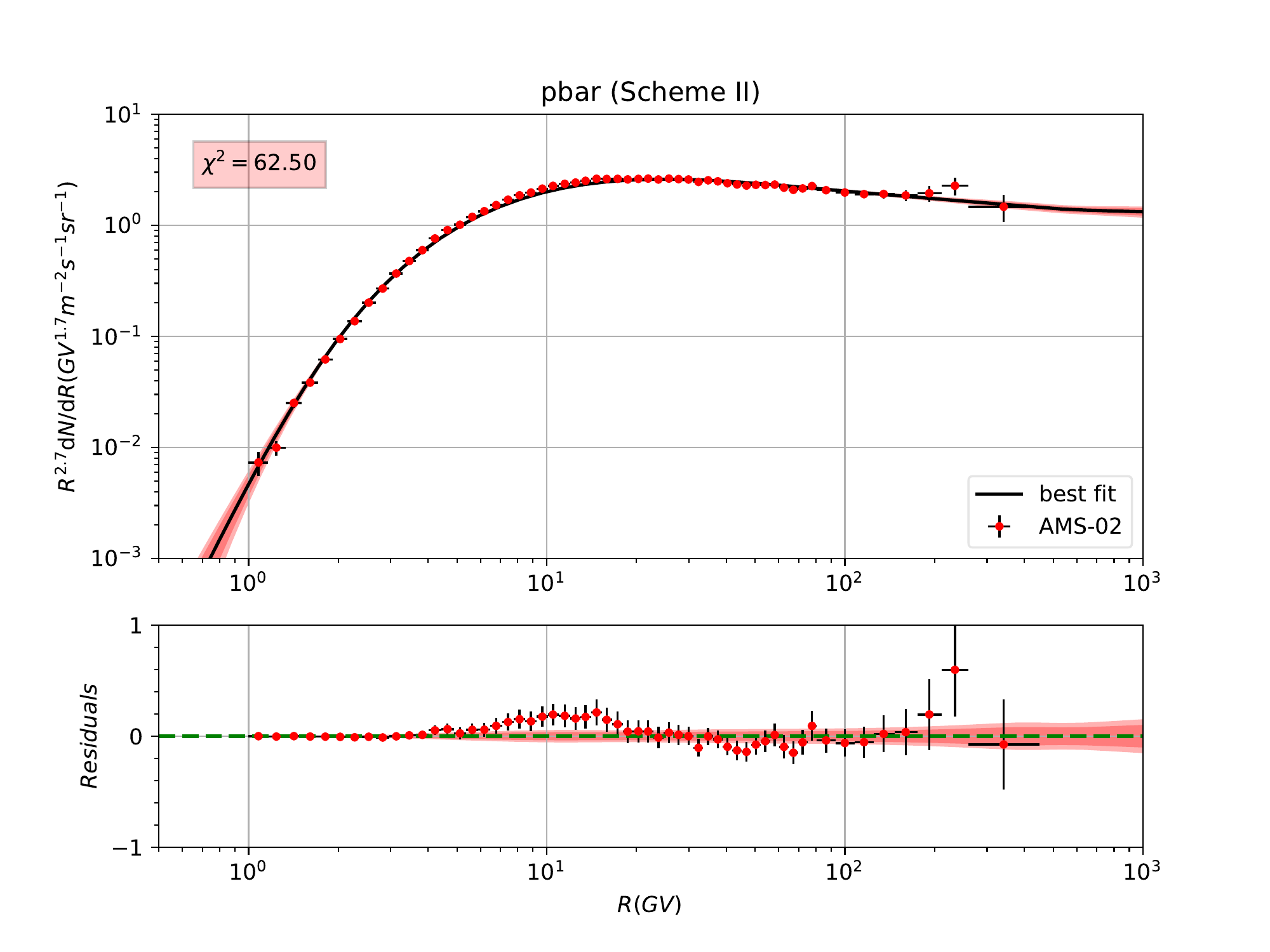}
  \includegraphics[width=0.43\textwidth]{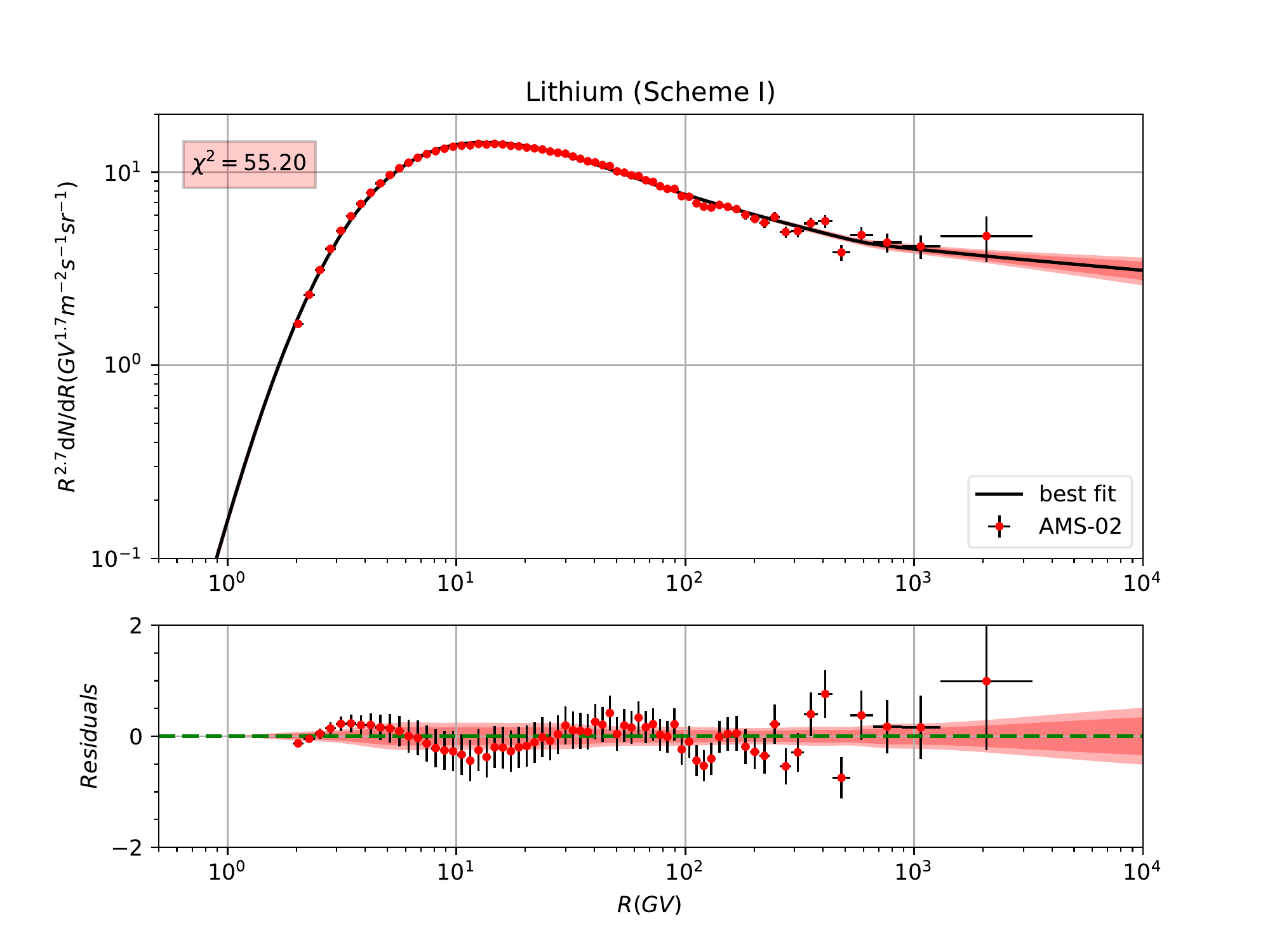}
  \includegraphics[width=0.43\textwidth]{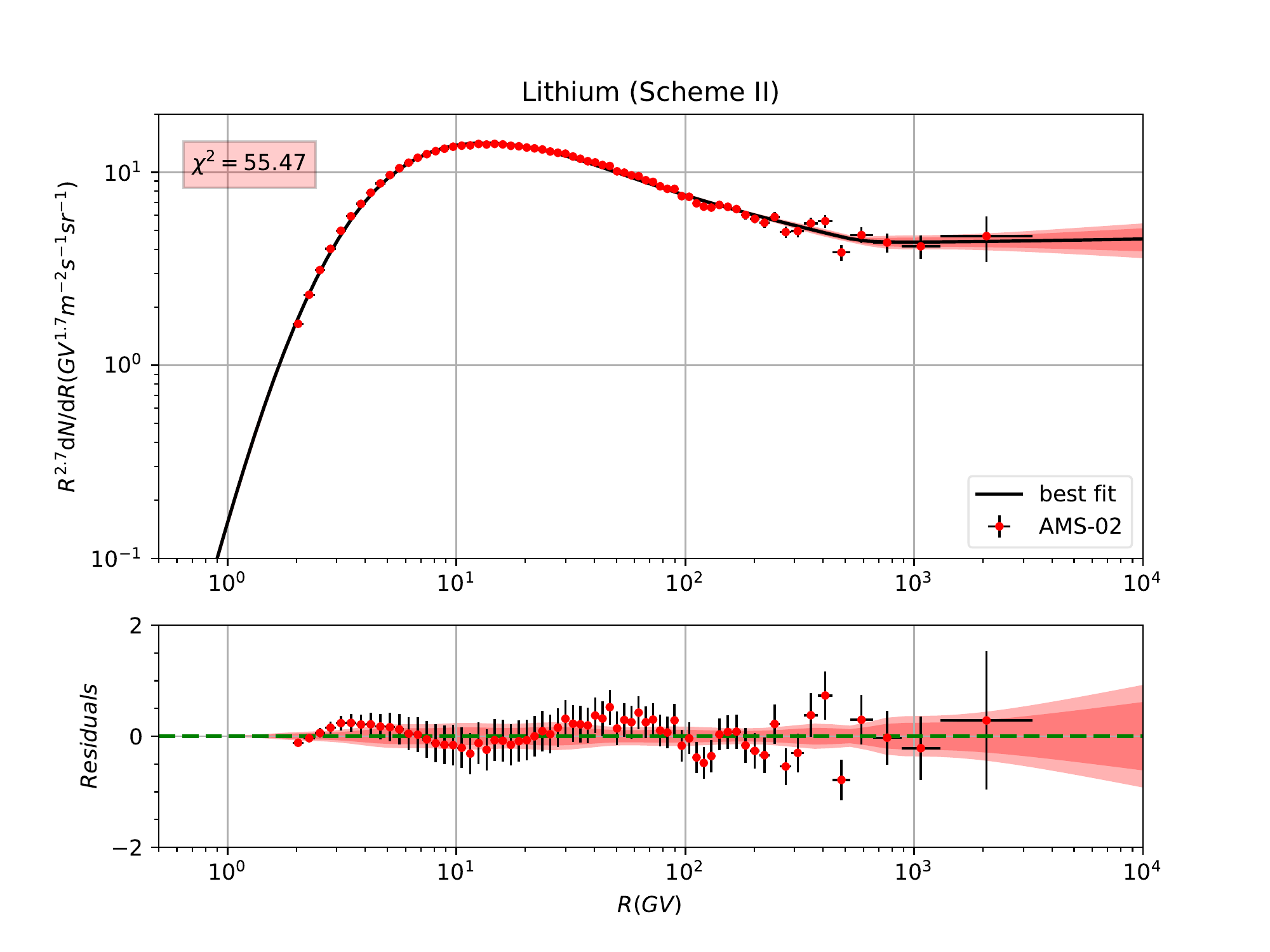}
  \includegraphics[width=0.43\textwidth]{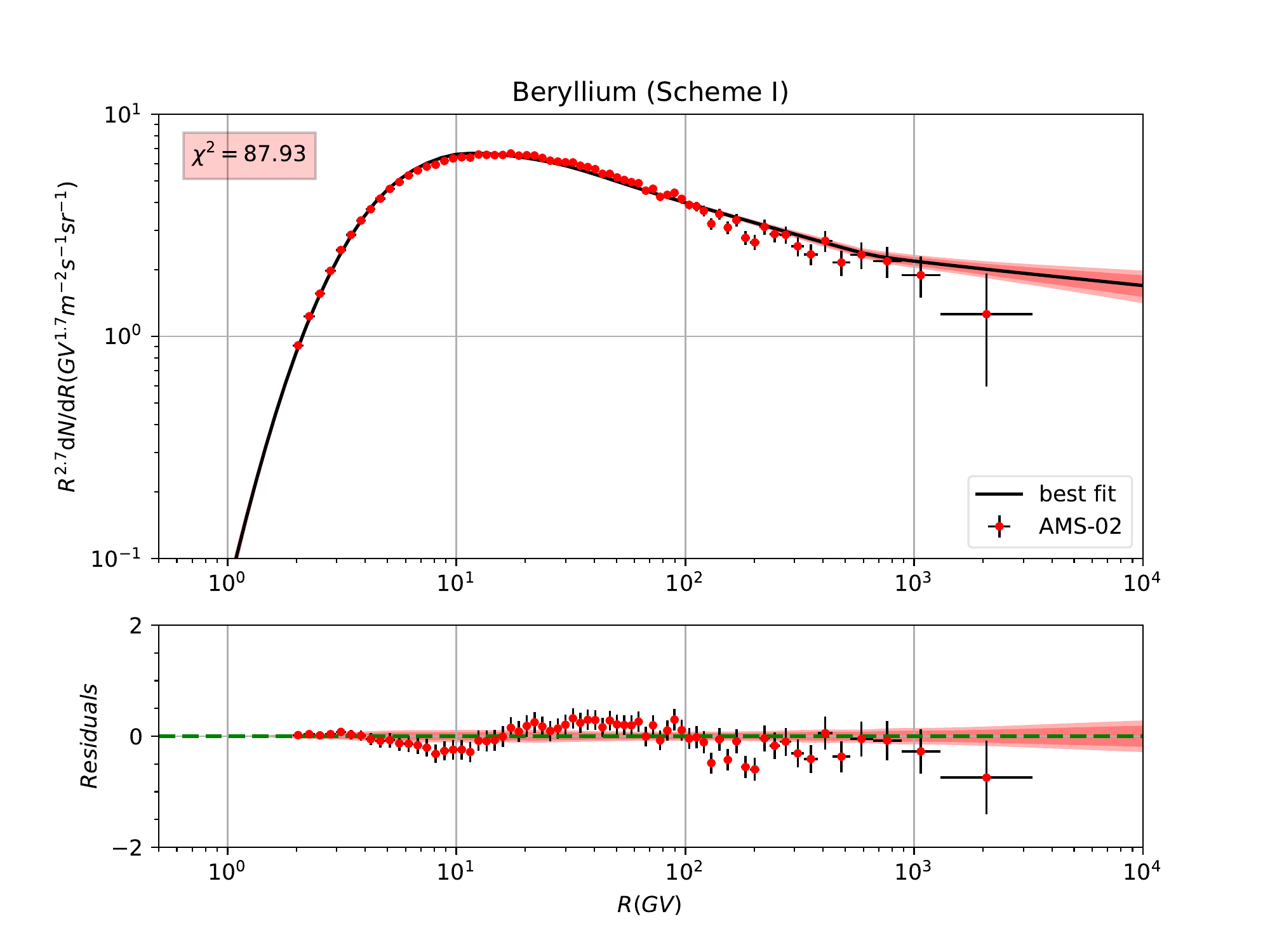}
  \includegraphics[width=0.43\textwidth]{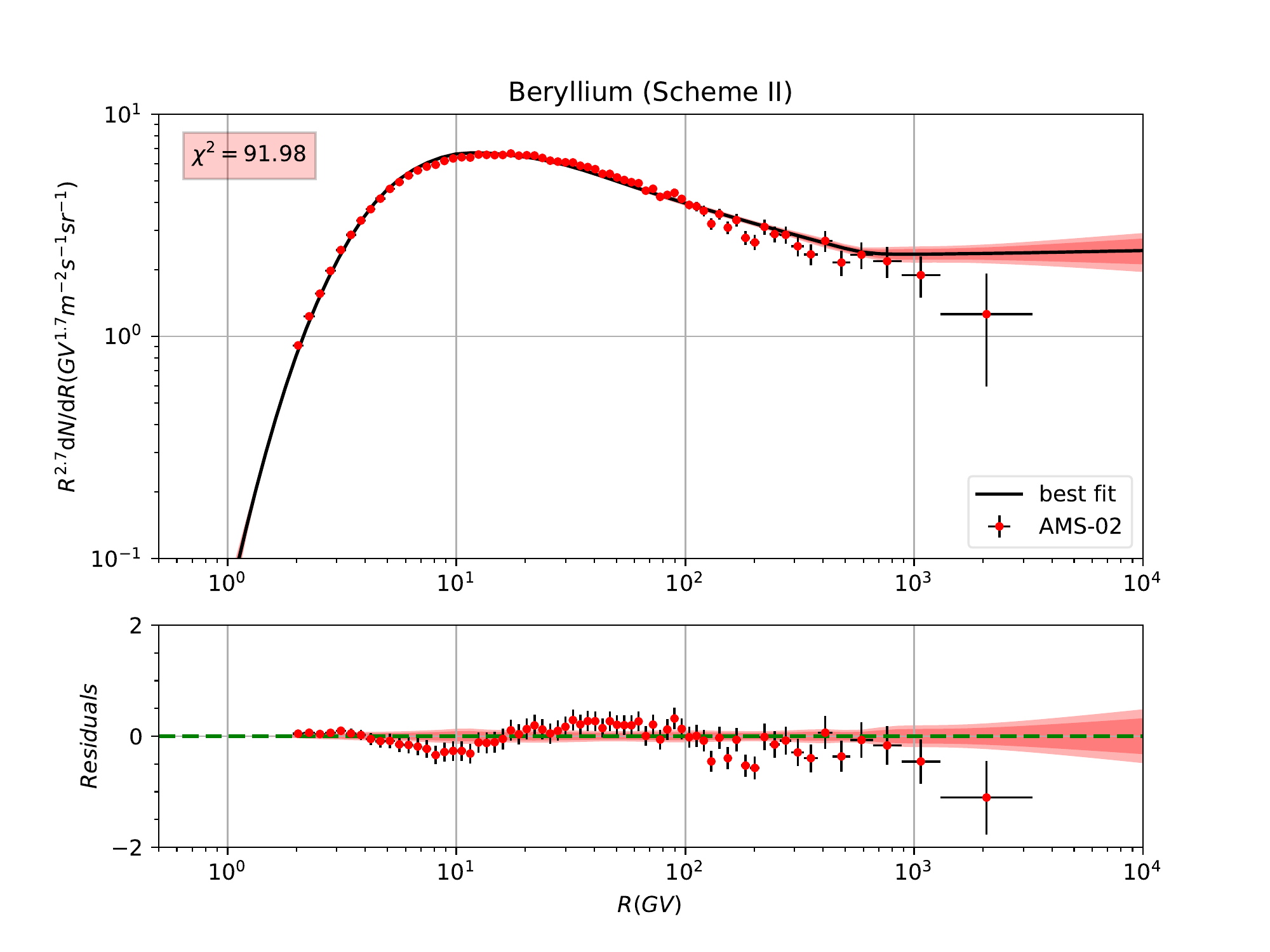}
  \includegraphics[width=0.43\textwidth]{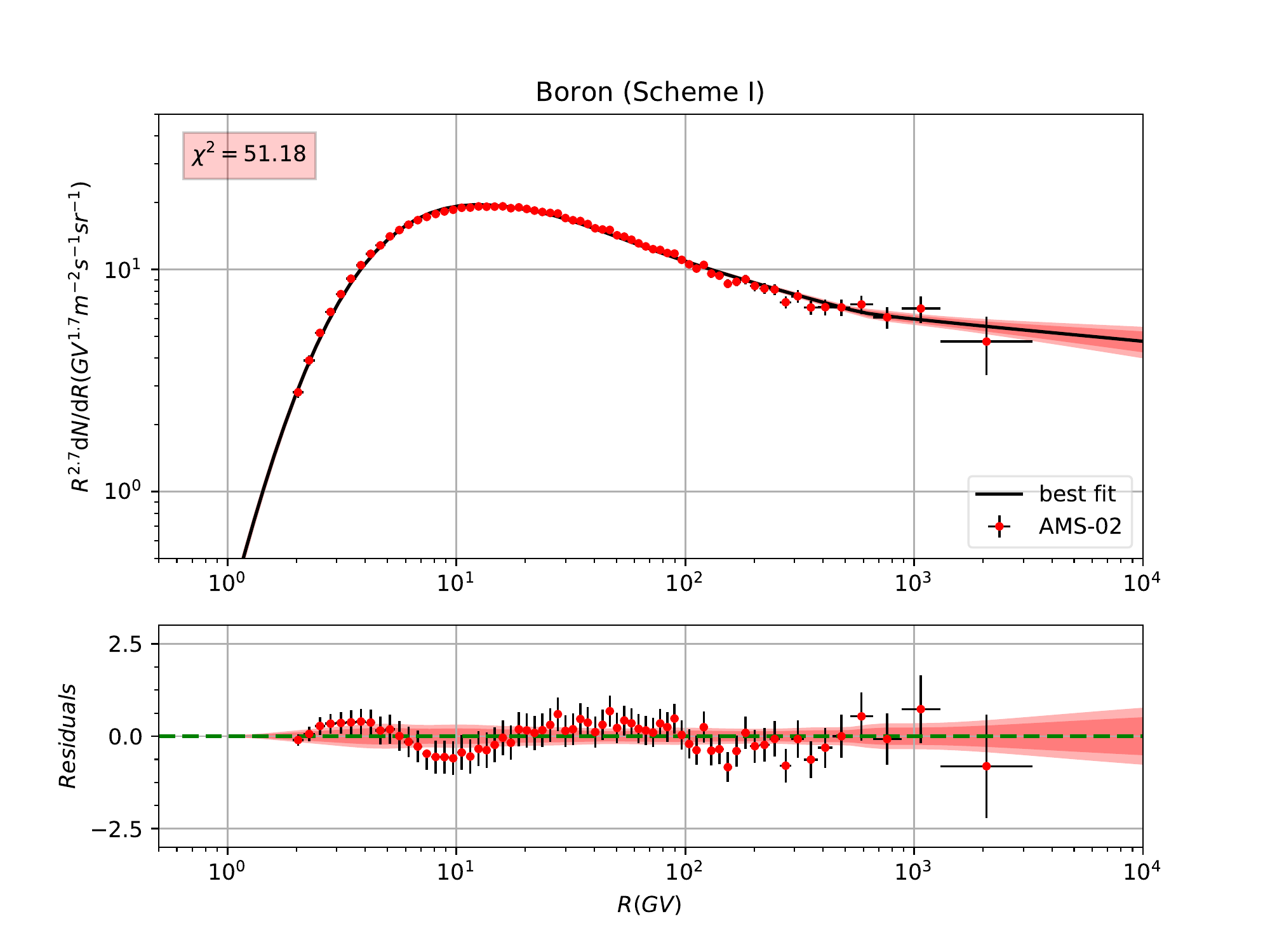}
  \includegraphics[width=0.43\textwidth]{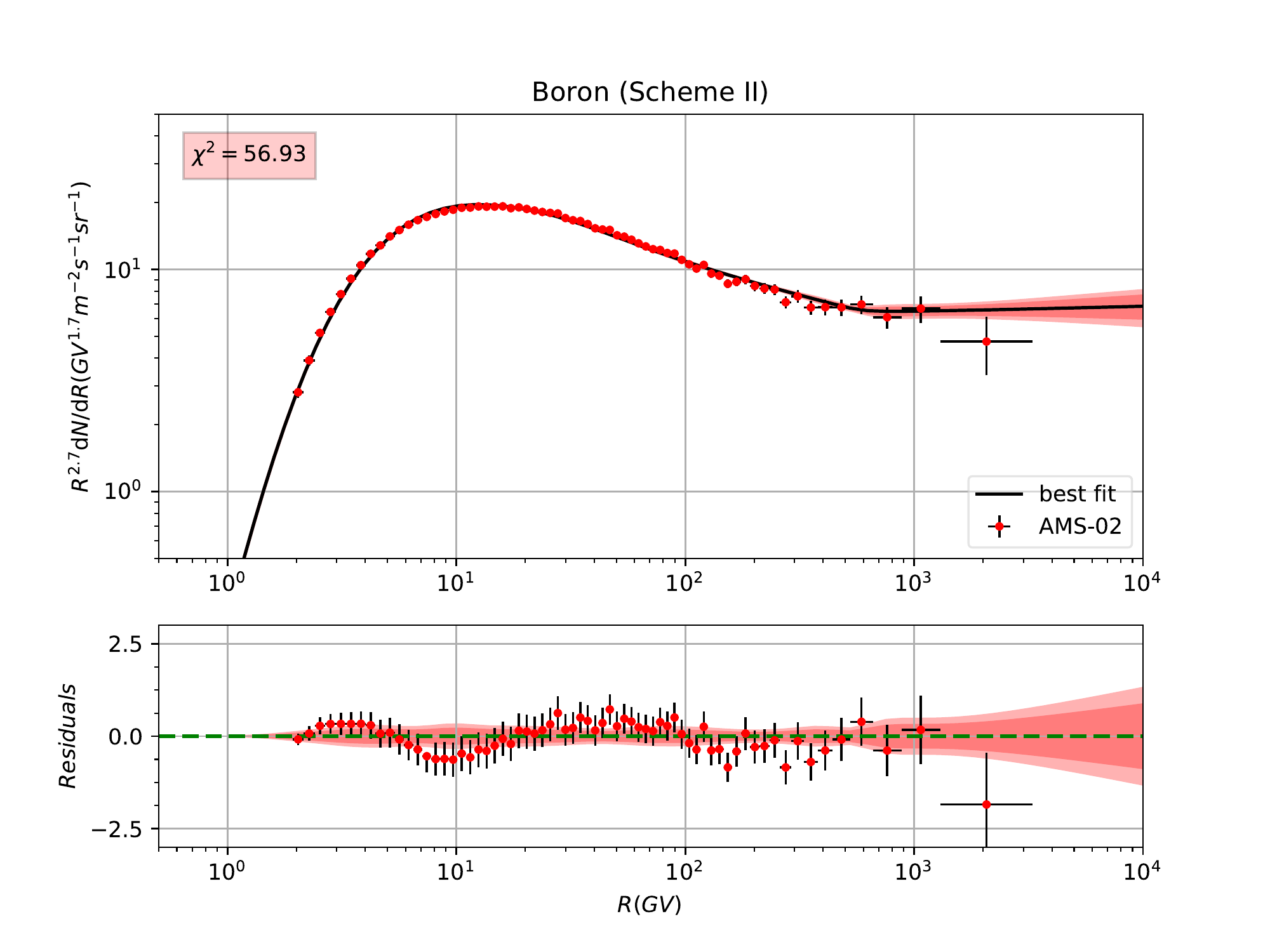}
  \caption{The global fitting results and the corresponding residuals to the secondary nuclei flux (anti-proton flux, lithium flux, beryllium flux and boron flux) for two schemes. The $2\sigma$ (deep red) and $3\sigma$ (light red) bounds are also shown in the figures. The relevant $\chi^{2}$ of each nuclei species is given in the sub-figures as well.}
\label{fig:secondary_results}
\end{figure*}

As in our previous works \citep{Niu2018,Niu2017_dampe1,Niu2017_dampe2}, we use the Markov Chain Monte Carlo (MCMC) algorithm  to  determine the posterior probability distribution of the parameters in Scheme I and II. We take the samples of the parameters as their posterior probability distribution function (PDF) after the Markov Chains have reached their equilibrium states. The best-fit results and the corresponding residuals of the primary nuclei flux for two schemes are given in Figure \ref{fig:primary_results}, and the corresponding results of the secondary nuclei flux are showed in Figure \ref{fig:secondary_results}. \footnote{Considering the correlations between different parameters, we could not get a reasonable reduced $\chi^{2}$ for each part of the data set independently. As a result, we present the $\chi^{2}$ for each part of the data set in Figures \ref{fig:primary_results}, \ref{fig:secondary_results}. }

The best-fit values, statistical mean values, standard deviations and allowed intervals at $95 \%$ CL for the parameters in $\boldsymbol{\theta}_{1}$ and $\boldsymbol{\theta}_{2}$  are shown in Table \ref{tab:scheme_params_I} and Table \ref{tab:scheme_params_II}, respectively. For best-fit results of the global fitting, we got $\chi^{2}/d.o.f = 383.45/521 $ for Scheme I and $\chi^{2}/d.o.f = 395.48/524 $ for Scheme II.

Generally speaking, the largest differences in the fitting results between Scheme I and II come from the fitting results of proton flux (which have a $\Delta \chi^{2} = \chi^{2}_{\mathrm{II, proton}} - \chi^{2}_{\mathrm{I, proton}} \simeq 12.6$) and helium flux (which have a $\Delta \chi^{2} = \chi^{2}_{\mathrm{I, He}} - \chi^{2}_{\mathrm{II, He}} \simeq 9.4$). This might come from that in Scheme I, we use independent breaks and slopes to describe the hardening in proton and other nuclei species. While in Scheme II, this lead to $ \chi^{2}_{\mathrm{I, proton}} < \chi^{2}_{\mathrm{II, proton}}$. At the same time, because the slopes in He, C and O spectra have almost the same value, it is natural that the $\delta_{2}$ would be mainly determined by these nuclei species in the global fitting, which leads to a result of  $\chi^{2}_{\mathrm{I, He}} > \chi^{2}_{\mathrm{II, He}}$.

In other cases, there are no obvious differences between the results of the two schemes. Moreover, we can see that in Figure \ref{fig:secondary_results}, the predicted tendency of the secondary nuclei spectra is different between the two schemes. Scheme I predicts a softer spectra when $R \gtrsim 1 \TV$ than Scheme II. This would be tested by high-rigidity ($> 1 \TV$) secondary nuclei data released in the future.

\begin{table*}[!htbp]
\caption{
Constraints on the parameters in set $\boldsymbol{\theta}_{1}$. The prior interval, best-fit value, statistic mean, standard deviation and the allowed range at $95\%$ CL are listed for parameters. With $\chi^{2}/d.o.f = 383.45/521 $ for best-fit result.}

\begin{center}
\begin{tabular}{lllll}
  \hline\hline
ID  &Prior & Best-fit &Posterior mean and   &Posterior 95\%    \\
    &range&value  &Standard deviation & range  \\
\hline
$D_{0}\ (10^{28}\cm^{2}\s^{-1})$ & [1, 30]    & 18.36 & 17.69$\pm$1.96 & [14.58, 20.97]\\

$\delta$                         & [0.1, 1.0] & 0.284 & 0.284$\pm$0.005& [0.277, 0.293]\\

$z_h\ (\kpc)$                    & [0.5, 30.0]& 11.30 & 10.32$\pm$1.43  & [7.85, 12.31]\\

$v_{A}\ (\km/\s)$                & [0, 80]    & 56.89 & 57.74$\pm$2.76 & [53.39, 62.31]\\

\hline

$R_{\p1}\ (\GV)$                 & [1, 30]    & 24.61 & 24.13$\pm$1.38 & [21.48, 26.28]\\

$R_{\p2}\ (\GV)$                 & [60, 1000] &528.31 &673.70$\pm$125.00 & [484.97, 893.13]\\

$\nu_{\p1}$                      & [1.0, 4.0] & 2.177 &2.172$\pm$0.015 & [2.145, 2.196]\\

$\nu_{\p2}$                      & [1.0, 4.0] & 2.474 &2.474$\pm$0.007 & [2.463, 2.485]\\

$\nu_{\p3}$                      & [1.0, 4.0] & 2.367 &2.352$\pm$0.015 & [2.326, 2.375]\\

$R_{\A1}\ (\GV)$                 & [1, 30]    & 22.23 & 21.47$\pm$1.06 & [19.91, 23.43]\\

$R_{\A2}\ (\GV)$                 & [60, 1000] & 540.03&504.07$\pm$68.20 & [400.28, 622.79]\\

$\nu_{\A1}$                      & [1.0, 4.0] & 2.096 &2.082$\pm$0.015 & [2.056, 2.107]\\

$\nu_{\A2}$                      & [1.0, 4.0] & 2.411 &2.409$\pm$0.006 & [2.401, 2.420]\\

$\nu_{\A3}$                      & [1.0, 4.0] & 2.252 &2.259$\pm$0.015 & [2.231, 2.283]\\

\hline

$N_{p}\ \footnote{Post-propagated normalization flux of protons at 100 GeV in unit $10^{-2}\m^{-2}\s^{-1}\sr^{-1}\GeV^{-1}$}$
                                 & [1, 8]     & 4.45 & 4.45$\pm$0.02   & [4.42, 4.48]\\

$\che$                           & [0.1, 5.0] & 0.643 & 0.645$\pm$0.004   & [0.638, 0.652]\\

$\cc$                            & [0.1, 5.0] & 0.551 & 0.553$\pm$0.005   & [0.545, 0.561]\\

$\co$                            & [0.1, 5.0] & 0.504 & 0.504$\pm$0.008   & [0.492, 0.518]\\

$\cpbar$                         & [0.1, 5.0] & 1.72 & 1.74$\pm$0.10   & [1.58, 1.91]\\

$\cli$                           & [0.1, 5.0] & 1.43 & 1.44$\pm$0.07   & [1.31, 1.57]\\

$\cbe$                           & [0.1, 5.0] & 1.70 & 1.72$\pm$0.09   & [1.57, 1.87]\\

$\cb$                            & [0.1, 5.0] & 1.10 & 1.11$\pm$0.05   & [1.03, 1.19]\\

\hline

$\phip\ (\GV)$                   & [0, 1.5]   & 0.70 & 0.70$\pm$0.02   & [0.66, 0.74]\\

$\phihe\ (\GV)$                  & [0, 1.5]   & 0.61 & 0.60$\pm$0.02   & [0.56, 0.64]\\

$\phic\ (\GV)$                   & [0, 1.5]   & 0.72 & 0.71$\pm$0.02   & [0.67, 0.75]\\

$\phio\ (\GV)$                   & [0, 1.5]   & 0.74 & 0.72$\pm$0.03   & [0.68, 0.76]\\

$\phipbar\ (\GV)$                & [0, 1.5]   & 0.008 & 0.02$\pm$0.02   & [0.001, 0.054]\\

$\phili\ (\GV)$                  & [0, 1.5]   & 0.65 & 0.62$\pm$0.04   & [0.56, 0.69]\\

$\phibe\ (\GV)$                  & [0, 1.5]   & 0.27 & 0.27$\pm$0.04   & [0.20, 0.33]\\

$\phib\ (\GV)$                   & [0, 1.5]   & 0.63 & 0.62$\pm$0.04   & [0.56, 0.69]\\

\hline\hline
\end{tabular}
\end{center}
\label{tab:scheme_params_I}
\end{table*}

\begin{table*}[!htbp]
\caption{
Constraints on the parameters in set $\boldsymbol{\theta}_{2}$. The prior interval, best-fit value, statistic mean, standard deviation and the allowed range at $95\%$ CL are listed for parameters. With $\chi^{2}/d.o.f = 395.48/524 $ for best-fit result.}
\begin{center}
\begin{tabular}{lllll}
  \hline\hline
ID  &Prior & Best-fit &Posterior mean and   &Posterior 95\%    \\
    &range&value  &Standard deviation & range  \\
\hline
$D_{0}\ (10^{28}\cm^{2}\s^{-1})$ & [1, 30]    & 18.27 & 17.98$\pm$1.22  & [15.98, 19.94]\\

$\Rbr\ (\GV)$                    & [200, 800] & 541.73& 559.80$\pm$73.35& [455.97, 693.90]\\

$\delta_{1}$                     & [0.1, 1.0] & 0.275 & 0.278$\pm$0.005 & [0.269, 0.287]\\

$\delta_{2}$                     & [0.1, 1.0] & 0.139 & 0.148$\pm$0.013 & [0.127, 0.170]\\

$z_h\ (\kpc)$                    & [0.5, 30.0]& 8.53 & 8.50$\pm$0.05   & [8.42, 8.58]\\

$v_{A}\ (\km/\s)$                & [0, 80]    & 65.66 & 65.04$\pm$3.80 & [59.01, 71.27]\\

\hline

$R_{\p}\ (\GV)$                  & [1, 30]    & 27.88 & 26.56$\pm$1.83 & [23.87, 29.38]\\

$\nu_{\p1}$                      &[1.0, 4.0]  & 2.203 & 2.191$\pm$0.016& [2.163, 2.217]\\

$\nu_{\p2}$                      &[1.0, 4.0]  & 2.494 & 2.487$\pm$0.007& [2.475, 2.498]\\

$R_{\A}\ (\GV)$                  & [1, 30]    & 20.71 & 20.84$\pm$0.79 & [19.63, 22.20]\\

$\nu_{\A1}$                      &[1.0, 4.0]  & 2.073 & 2.066$\pm$0.014  & [2.042, 2.087]\\

$\nu_{\A2}$                      &[1.0, 4.0]  & 2.407 & 2.406$\pm$0.006  & [2.397, 2.416]\\

\hline

$N_{p}\ \footnote{Post-propagated normalization flux of protons at 100 GeV in unit $10^{-2}\m^{-2}\s^{-1}\sr^{-1}\GeV^{-1}$}$
                                 & [1, 8]     & 4.49  &4.47$\pm$0.02   & [4.43, 4.51]\\

$\che$                           & [0.1, 5.0] & 0.644  &0.646$\pm$0.004   & [0.639, 0.653]\\

$\cc$                            & [0.1, 5.0] & 0.551  &0.552$\pm$0.005   & [0.545, 0.560]\\

$\co$                            & [0.1, 5.0] & 0.497  &0.500$\pm$0.007   & [0.489, 0.512]\\

$\cpbar$                         & [0.1, 5.0] & 1.89  &1.89$\pm$0.11   & [1.70, 2.07]\\

$\cli$                           & [0.1, 5.0] & 1.53  &1.52$\pm$0.08   & [1.39, 1.65]\\

$\cbe$                           & [0.1, 5.0] & 1.80  &1.79$\pm$0.08   & [1.65, 1.92]\\

$\cb$                            & [0.1, 5.0] & 1.18  &1.17$\pm$0.06   & [1.08, 1.26]\\

\hline

$\phip\ (\GV)$                   & [0, 1.5]   & 0.73  &0.72$\pm$0.03   & [0.67, 0.76]\\

$\phihe\ (\GV)$                  & [0, 1.5]   & 0.57  &0.56$\pm$0.02   & [0.52, 0.60]\\

$\phic\ (\GV)$                   & [0, 1.5]   & 0.69  &0.67$\pm$0.03   & [0.63, 0.72]\\

$\phio\ (\GV)$                   & [0, 1.5]   & 0.71  &0.70$\pm$0.03   & [0.65, 0.74]\\

$\phipbar\ (\GV)$                & [0, 1.5]   & 0.002&0.01$\pm$0.01   & [0.0006, 0.0352]\\

$\phili\ (\GV)$                  & [0, 1.5]   & 0.56  &0.55$\pm$0.04   & [0.48, 0.61]\\

$\phibe\ (\GV)$                  & [0, 1.5]   & 0.18  &0.17$\pm$0.04   & [0.10, 0.24]\\

$\phib\ (\GV)$                   & [0, 1.5]   & 0.56  &0.56$\pm$0.04   & [0.49, 0.62]\\

\hline\hline
\end{tabular}
\end{center}
\label{tab:scheme_params_II}
\end{table*}

\subsection{Propagation Parameters}

\begin{figure}[!htbp]
  \centering
  \includegraphics[width=0.46\textwidth]{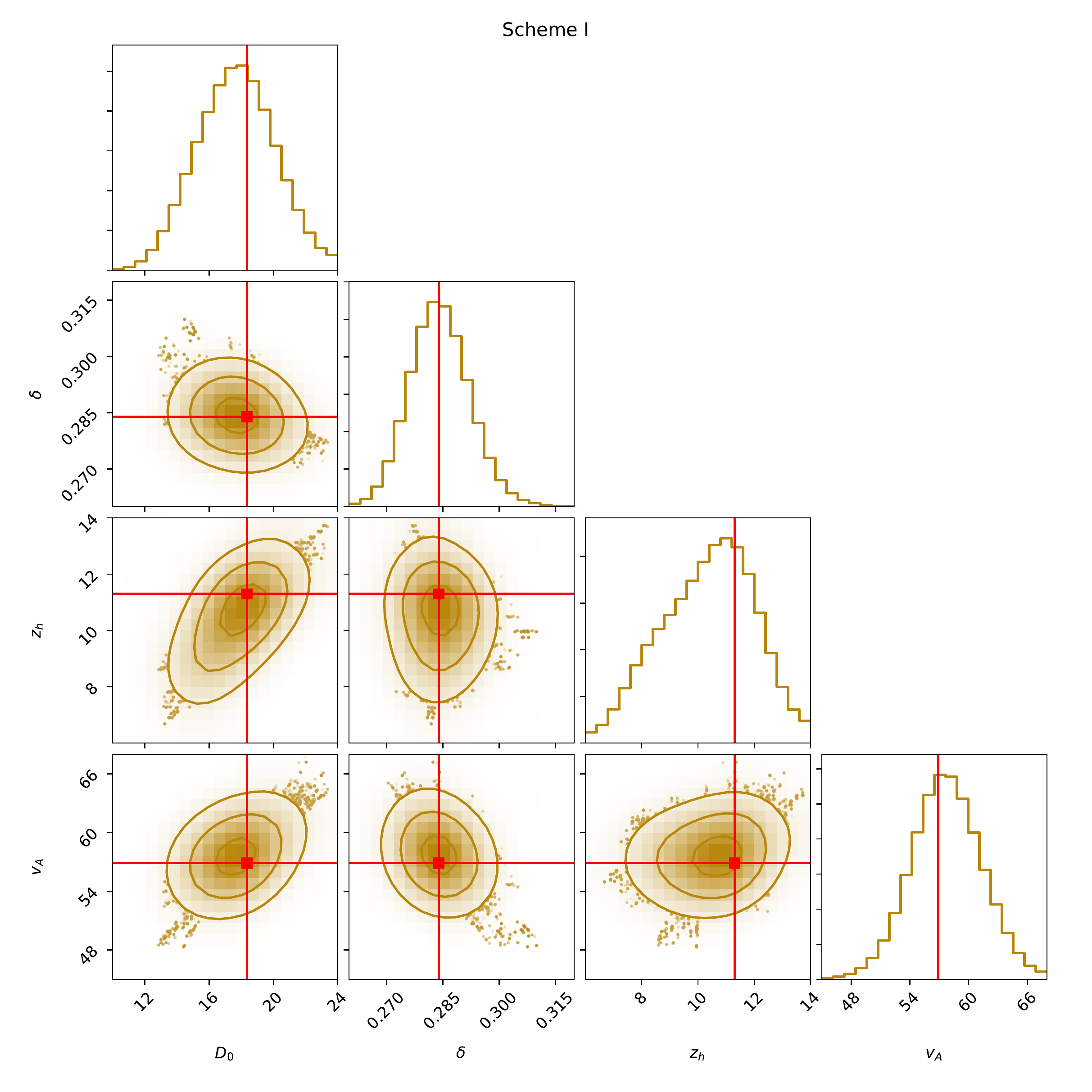}
  \caption{Fitting 1D probability and 2D credible regions of posterior PDFs for the combinations of all propagation parameters from Scheme I. The regions enclosing $\sigma$, $2\sigma$ and $3\sigma$ CL are shown in step by step lighter golden. The red cross lines and marks in each plot indicate the best-fit value (largest likelihood).}
\label{fig:prop_params_sch1}
\end{figure}

\begin{figure}[!htbp]
  \centering
  \includegraphics[width=0.48\textwidth]{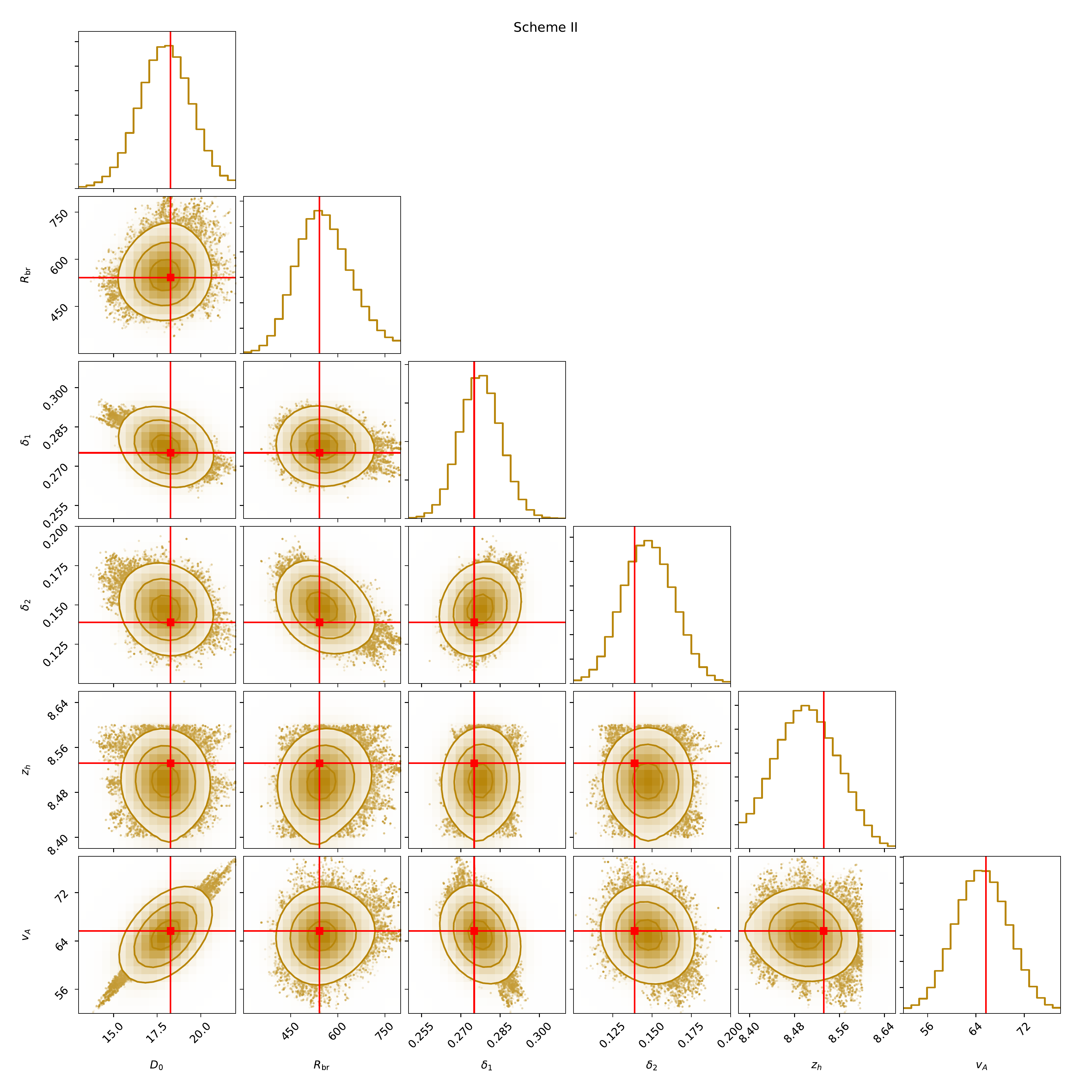}
  \caption{Same as Figure \ref{fig:prop_params_sch1}, but for Scheme II.}
\label{fig:prop_params_sch2}
\end{figure}

The results of posterior probability distributions of the propagation parameters are shown in  Figure \ref{fig:prop_params_sch1} (Scheme I), and Figure \ref{fig:prop_params_sch2} (Scheme II).

The most obvious differences between the fitting results of the parameters in this work and some previous works (see, e.g., \citet{Yuan2017,Niu2018}) are the values of $D_{0}$, $z_{h}$, $\delta$, and $v_{A}$. Compared with our previous work \citep{Niu2018},  $D_{0}$, $z_{h}$, and $v_{A}$ have relatively larger values here (especially in Scheme I), while $\delta$ have a smaller value in this work. In view of the data sets and parameters configuration (the hardening of the spectra have been fully considered), the results in this work should have a higher level of confidence.

In Scheme I, the $\delta$ value obtained is obviously smaller than that in our previous work (in which $\delta \simeq 3.5 - 3.7 $) \citep{Niu2018}. This is because the added breaks in the primary source injection of proton ($\sim 480 - 890 \GV$) and other primary nuclei species ($\sim 400 - 620 \GV$) could take charge of the observed hardening in their observed spectra, rather than using only one break in the source injection and letting the only $\delta$ compromise the different slopes in high energy regions in \citet{Niu2018}.

In Scheme II, we got a high-rigidity break at $\sim (450 - 700) \GV$ in the diffusion coefficient, and a slope $\delta_{2} \sim (0.13 - 0.17)$ above the break. Although the value of $\delta_{1}$ is not in the same posterior distribution region as that in \citet{Genolini2017} (which gave $\delta_{1} \sim 0.5 - 0.7$), we got a similar value of $\Delta \delta = \delta_{1} - \delta_{2} \sim 0.14 $. Considering the simplifications in \citet{Genolini2017} to do calculation for catching the key points in their work, we could conclude that we get a consistent result compared with their work.

Moreover, whether a high-rigidity break in diffusion coefficient is needed in current AMS-02 nuclei data if we have already considered the high-rigidity break in primary source injection? Some conclusions are proposed in \citet{Yuan2018}. It is $\delta$ value dependent. When $\delta \sim 0.3$, it is not needed such a break; while $\delta \sim 0.5 - 0.7$, such a break is needed to reproduce current spectra data.

Another point should be noted is that in Scheme II, the  uncertainty of $z_{h}$ could reach down to 0.05, which might be caused by the special configurations in this scheme (employing a break in diffusion coefficient to account for the hardening of all the primary and secondary spectra and not considering the differences between the spectra of proton and other species).

\subsection{Primary Source Injection Parameters}

\begin{figure*}[!htbp]
  \centering
    \includegraphics[height=0.95\textheight,width=0.97\textwidth]{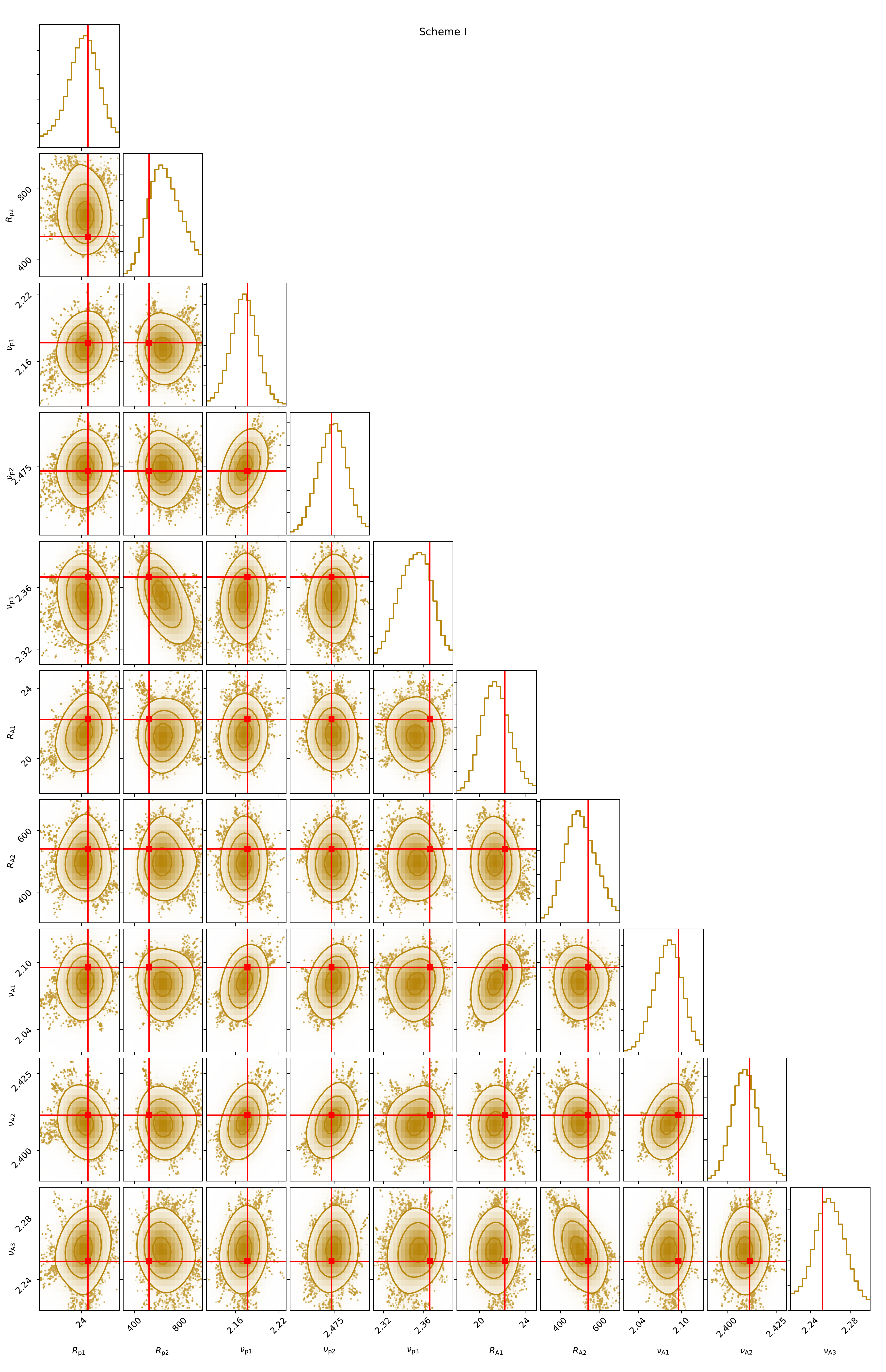}
  \caption{Fitting 1D probability and 2D credible regions of posterior PDFs for the combinations of all primary source injection parameters from Scheme I. The regions enclosing $\sigma$, $2\sigma$ and $3\sigma$ CL are shown in step by step lighter golden. The red cross lines and marks in each plot indicate the best-fit value (largest likelihood).}
\label{fig:inj_params_sch1}
\end{figure*}

\begin{figure*}[!htbp]
  \centering
  \includegraphics[width=0.96\textwidth]{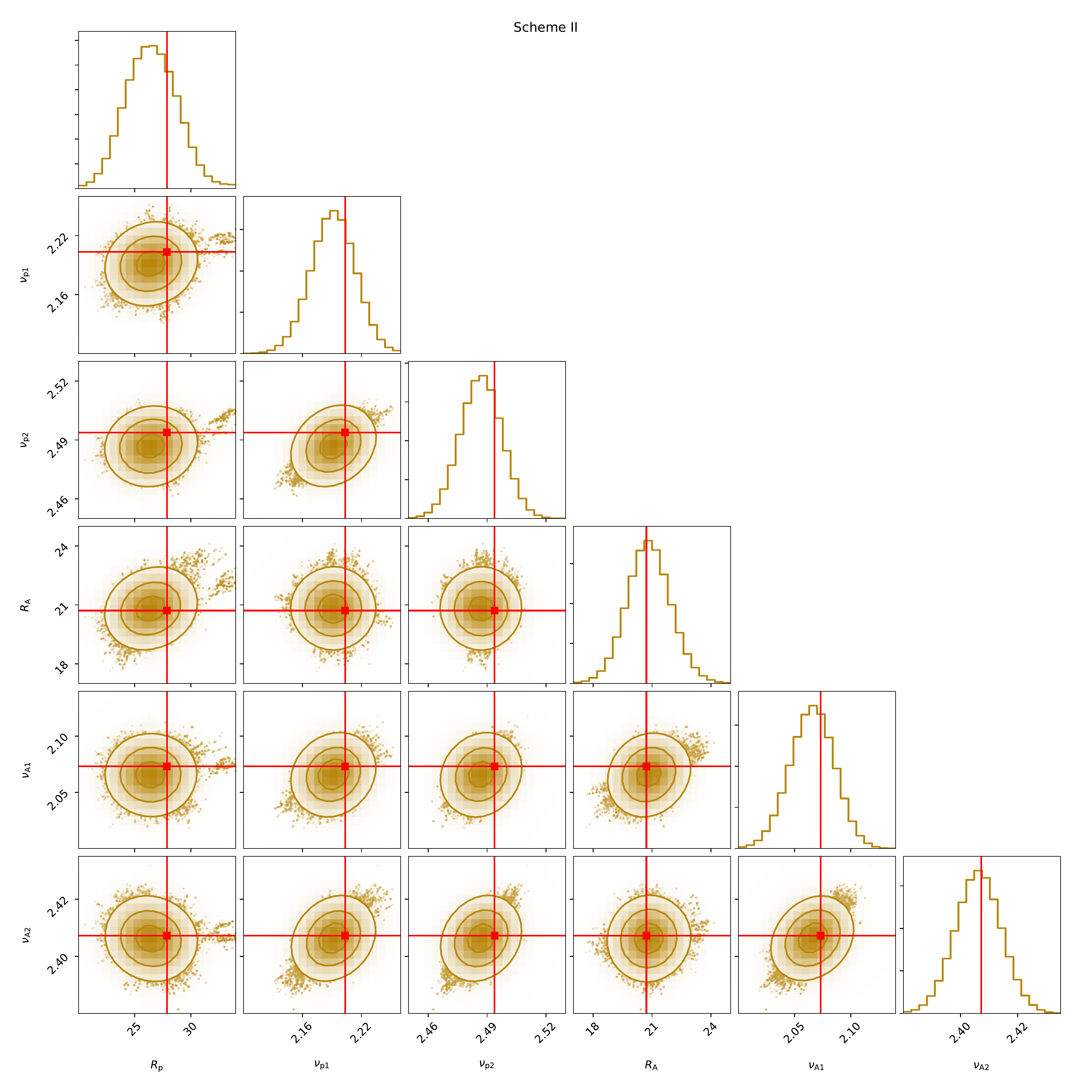}
  \caption{Same as Figure \ref{fig:inj_params_sch1}, but for Scheme II.}
\label{fig:inj_params_sch2}
\end{figure*}

The results of posterior probability distributions of the primary source injection parameters are presented in  Figure \ref{fig:inj_params_sch1} (Scheme I), and Figure \ref{fig:inj_params_sch2} (Scheme II).

Same as our previous works \citep{Niu2018,Niu2017_dampe1}, the rigidity breaks and slopes are obviously different between proton and other nuclei species in both schemes.  Particularly, in Scheme I, the differences between the primary source injection high-rigidity slopes have values of  $\nu_{\p2} - \nu_{\p3} \sim 0.1 $ (for proton) and $\nu_{\A2} - \nu_{\A3} \sim 0.15$ (for other nuclei species). This indicates that if we want to ascribe the hardening of the spectra to the primary source injections, the acceleration mechanisms in this energy region ($500 - 800 \GV$)  should be different between proton and other nuclei species.

In Scheme II, the hardening of the spectra is accounted by the break in diffusion coefficient, and the fitting results of the  breaks and slopes in primary source injections are consistent with that in Scheme I. Consequently, these fitted values are reliable.

\subsection{Normalization Parameters}

\begin{figure*}[!htbp]
  \centering
  \includegraphics[width=0.96\textwidth]{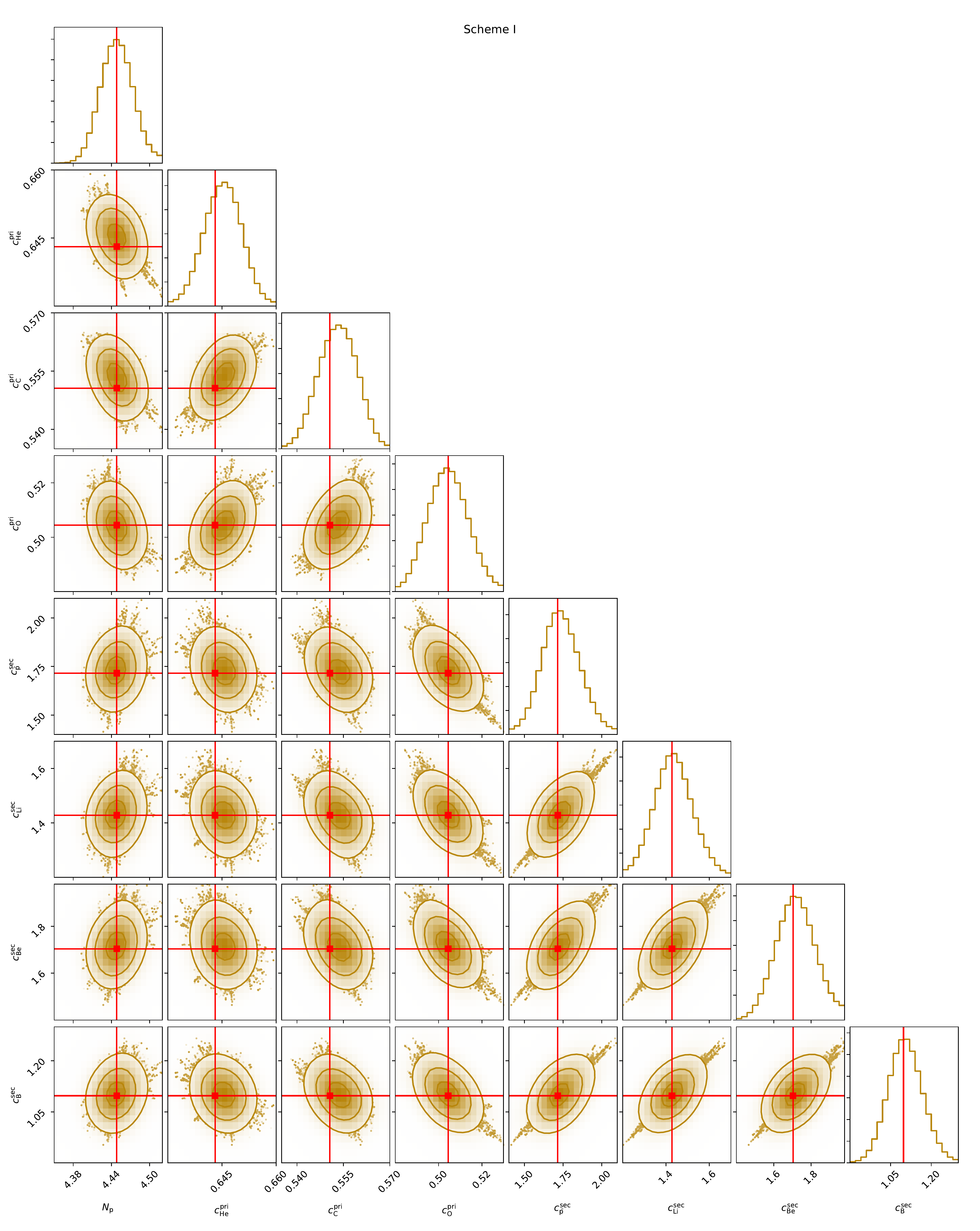}
  \caption{Fitting 1D probability and 2D credible regions of posterior PDFs for the combinations of all normalization parameters from Scheme I. The regions enclosing $\sigma$, $2\sigma$ and $3\sigma$ CL are shown in step by step lighter golden. The red cross lines and marks in each plot indicate the best-fit value (largest likelihood).}
\label{fig:c_params_sch1}
\end{figure*}

\begin{figure*}[!htbp]
  \centering
  \includegraphics[width=0.96\textwidth]{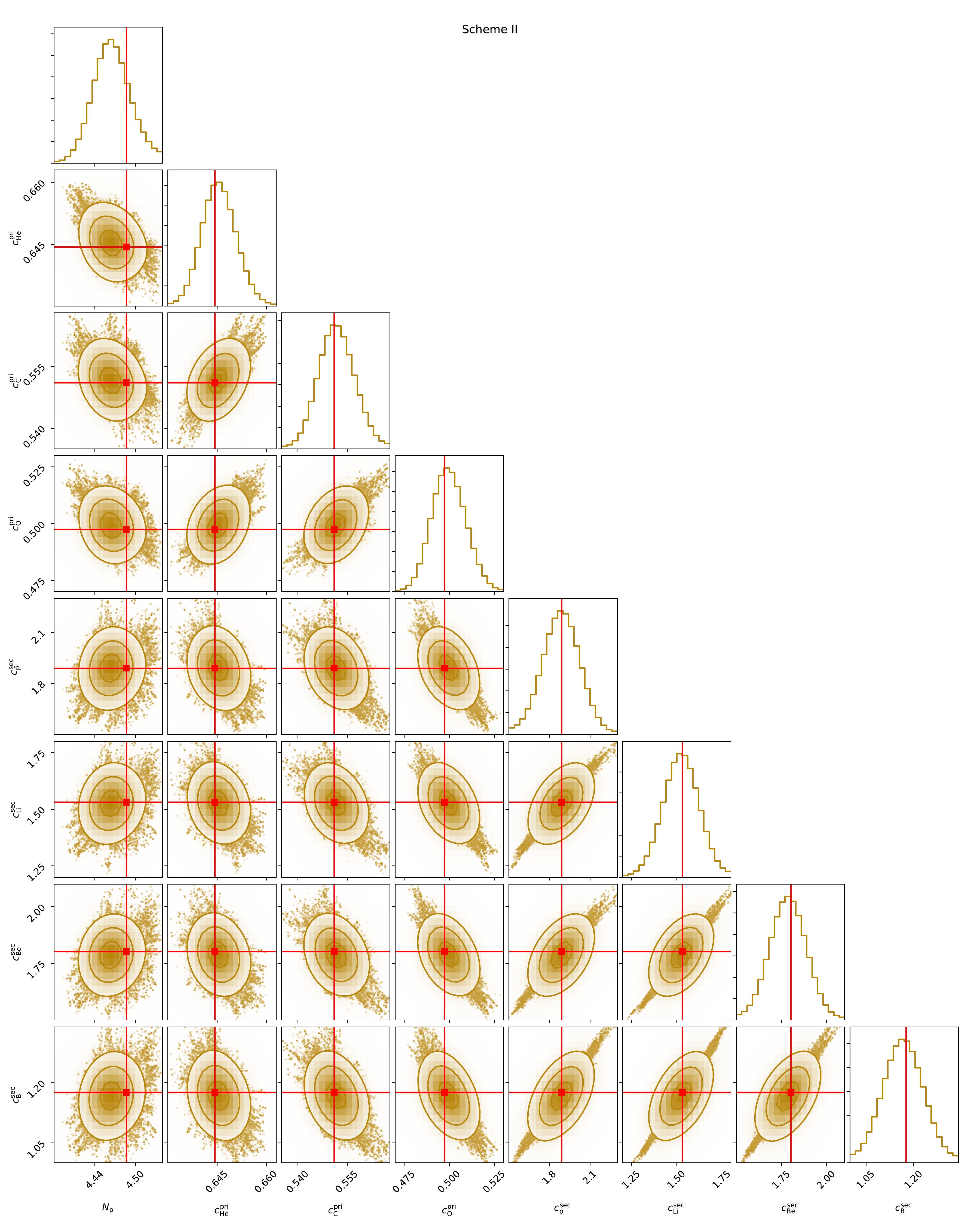}
  \caption{Same as Figure \ref{fig:c_params_sch1}, but for Scheme II.}
\label{fig:c_params_sch2}
\end{figure*}

The results of posterior probability distributions of the normalization parameters are given in  Figure \ref{fig:c_params_sch1} (Scheme I), and Figure \ref{fig:c_params_sch2} (Scheme II).

In Tables \ref{tab:scheme_params_I} and \ref{tab:scheme_params_II}, we find that the normalization parameters of the primary nuclei species have an uncertainty of $< 1\%$, and that of the secondary nuclei species $\sim 5\%$. This shows us the necessity to employ them in the global fitting. Although their fitted values have slight differences between Schemes I and II, the relative relations can be kept in both of the 2 schemes.

Interestingly, the re-scale factors of all the  primary nuclei species are $< 1.0$, and all that of  the secondary species are $> 1.0$.  Here, $\cpbar$ and $\cbe$ should be given more attentions because of the large deviations compared to other secondary species.

The value of $\cpbar$ in this work ($\sim $ 1.7 - 1.9) is obviously different from that in previous works ($\sim 1.3 - 1.4$),  which could be partially explained by the usage of an independent $\phipbar$ to modulate the low-rigidity data (which contributes corrections on $\cpbar$) in this work. At the same time, we should note that in Figure \ref{fig:phi_params_sch1} and \ref{fig:phi_params_sch2}, the PDFs of $\phipbar$ reveal that even an independent solar modulation potential $\phipbar$ (which is based on force-field approximation) could not handle the solar modulation effects on $\pbar$ which have a negative charge. In this view, $\cpbar$ could not be considered as a pure value to describe the uncertainty of $\pbar$ production cross section.

Besides $\cpbar$, the following large deviation value is $\cbe$. Not like $\pbar$, the solar modulation of Be could be well modeled in our fitting (see in Figures \ref{fig:phi_params_sch1} and \ref{fig:phi_params_sch2}). We cannot find any other reasons to interpret its specificity tentatively, and this needs more attention in future research.

\subsection{Solar Modulation Potentials}

\begin{figure*}[!htbp]
  \centering
  \includegraphics[width=0.96\textwidth]{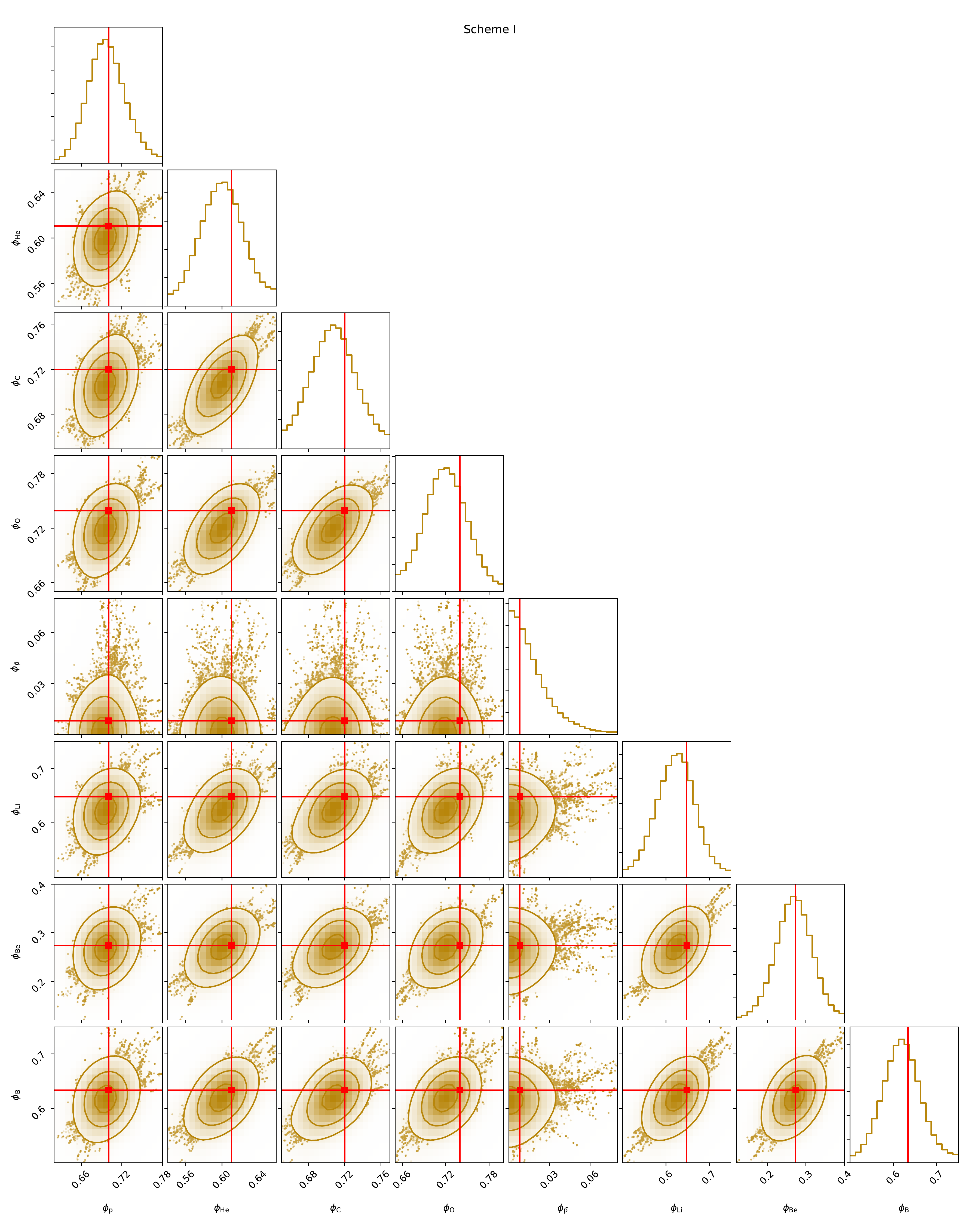}
  \caption{Fitting 1D probability and 2D credible regions of posterior PDFs for the combinations of all solar modulation potentials from Scheme I. The regions enclosing $\sigma$, $2\sigma$ and $3\sigma$ CL are shown in step by step lighter golden. The red cross lines and marks in each plot indicate the best-fit value (largest likelihood).}
\label{fig:phi_params_sch1}
\end{figure*}

\begin{figure*}[!htbp]
  \centering
  \includegraphics[width=0.96\textwidth]{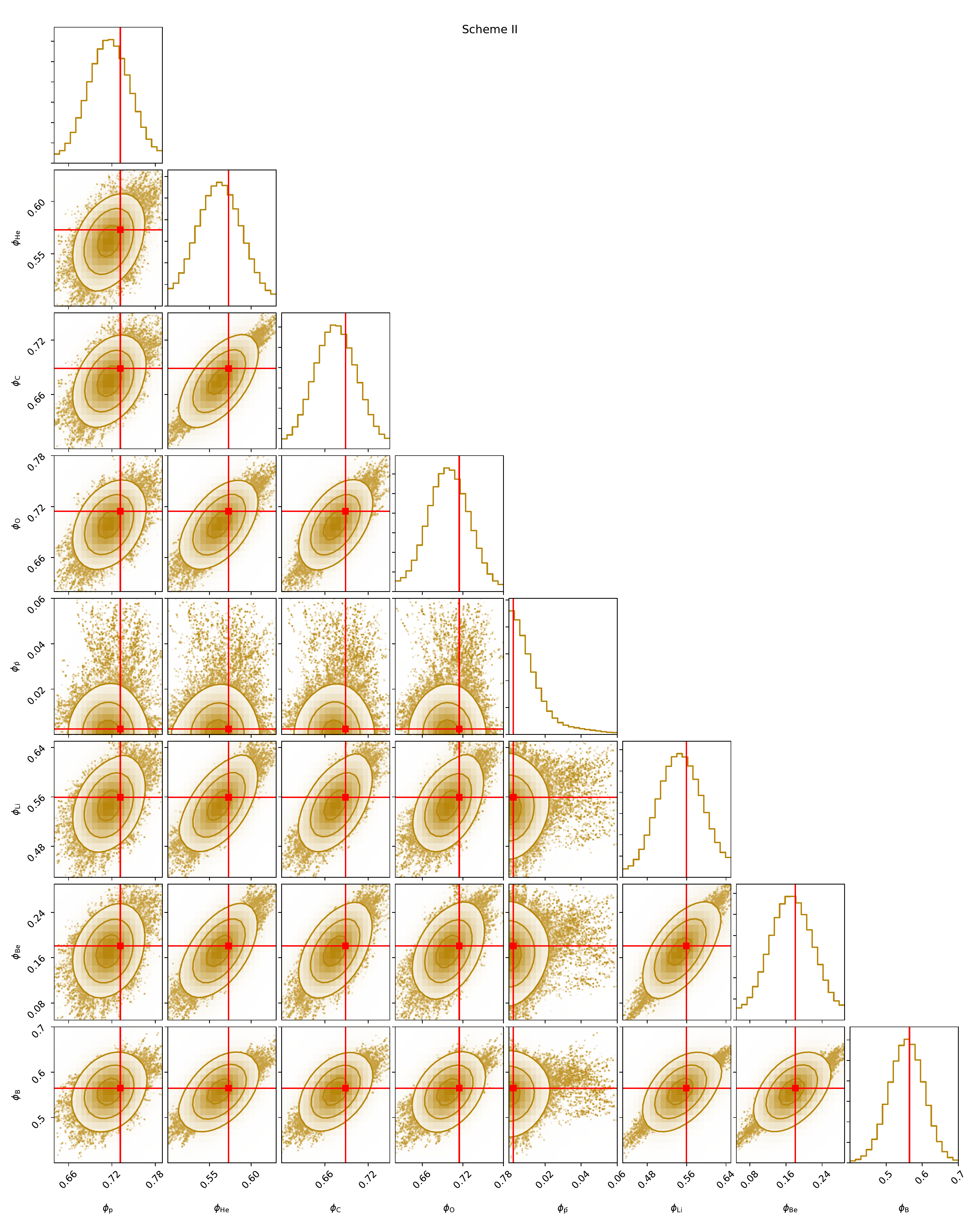}
  \caption{Same as Figure \ref{fig:phi_params_sch1}, but for Scheme II.}
\label{fig:phi_params_sch2}
\end{figure*}

The results of posterior probability distributions of the solar modulation potentials are shown in  Figure \ref{fig:phi_params_sch1} (Scheme I), and Figure \ref{fig:phi_params_sch2} (Scheme II). For convenience, the boxplot of all the $\phi_{i}$s in Scheme I and II are shown in Figure \ref{fig:phi_box}.

What is interesting is that, all these $\phi_{i}$s have almost similar values ($0.5 - 0.7 \GV$) except $\phipbar$ and $\phibe$. As we know, as an effective tool to handle solar modulation, force-field approximation is charge independent. The fitted values of $\phipbar$ clearly show that this approximation cannot deal with the solar modulation effects on $\pbar$ at current data levels. Removing $\phipbar$ not to talk, it is strange that $\phibe$ have a larger deviation compared with other $\phi_{i}$s. Considering the uncertainties and PDFs in Figures \ref{fig:phi_params_sch1} and \ref{fig:phi_params_sch2}, the values of $\phibe$ should be regarded seriously. The reasons and relevant physics behind $\phibe$ should be studied in further research.

\section{Nitrogen Spectrum as a Test}
\label{sec:test}

Nitrogen nuclei in CRs are thought to be produced both in astrophysical sources (mostly via the CNO cycle in stars \citep{Bethe1939,Chiappini2003,Henry2000}), and by the collisions of heavier nuclei with the ISM \citep{Strong2007,Blasi2013,Grenier2015}. As a result, the nitrogen spectrum is expected to contain both primary and secondary components, which is the ideal data set to test not only the propagation model, but also the primary source injections.

Recently released nitrogen spectrum from AMS-02 \citep{AMS02_N} with rigidity from 2.2 GV to 3.3 TV is used to do a test of the best-fit results in Section \ref{sec:fitting_results}. In the test, based on the best-fit results of Scheme I and II in Section \ref{sec:fitting_results}, the re-scale parameters of the primary and secondary components of nitrogen nuclei ($\cnp$ and $\cns$)\footnote{The relative abundance of nitrogen-14 has a default value of $1.828 \times 10^{2}$ in {\sc galprop}.}, and the solar modulation potential $\phin$ are set to be free parameters to do a global fitting on the nitrogen spectrum.

\begin{figure*}[!htbp]
  \centering
  \includegraphics[width=0.48\textwidth]{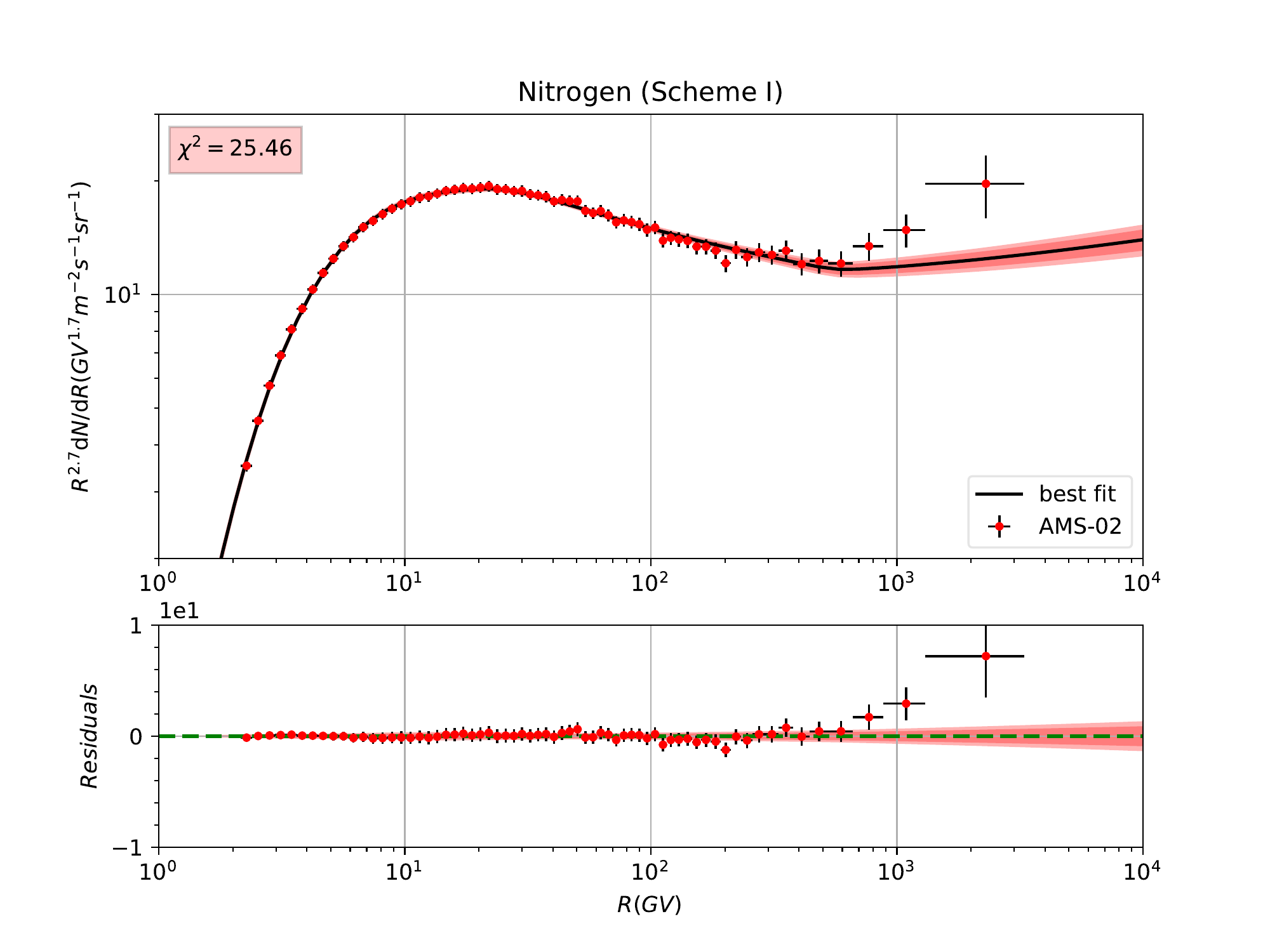}
  \includegraphics[width=0.48\textwidth]{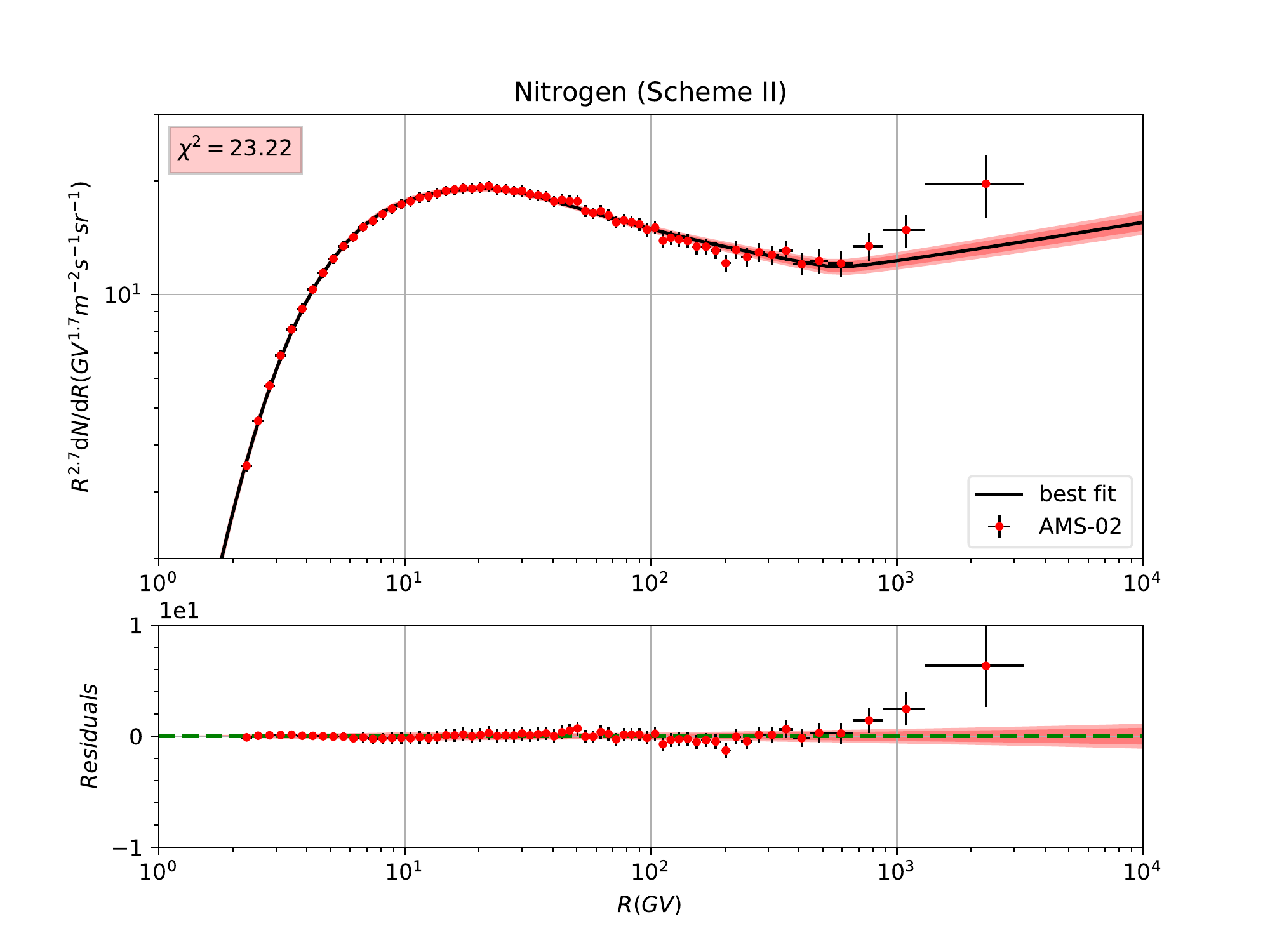}
  \caption{The global fitting results and the corresponding residuals to the nitrogen flux for two schemes. The $2\sigma$ (deep red) and $3\sigma$ (light red) bounds are also shown in the figures. The relevant $\chi^{2}$ of each nuclei species is given in the sub-figures as well.}
\label{fig:nitrogen}
\end{figure*}

As that have been done in Section \ref{sec:fitting_results}, MCMC algorithm is used to determine the best-fit results on the nitrogen spectrum (see in Figure \ref{fig:nitrogen}) and the constraints on $\cnp$, $\cns$, and $\phin$ (see in Tables \ref{tab:nitrogen_scheme_I} and \ref{tab:nitrogen_scheme_II}) of the 2 schemes.

Considering the same degree of freedom in 2 schemes in the test, the results of the 2 schemes could be compared directly. As it has been shown in Tables \ref{tab:nitrogen_scheme_I} and \ref{tab:nitrogen_scheme_II}, both schemes give us a quit good best-fit result ($\chi^{2}_{I,\mathrm{N}}/d.o.f. = 25.46/63$ and $\chi^{2}_{II,\mathrm{N}}/d.o.f. = 23.22/63$), except the last 3 points with large uncertainties (see in Figure \ref{fig:nitrogen}). Although $\chi^{2}_{II,\mathrm{N}}$ is slightly smaller than $\chi^{2}_{I,\mathrm{N}}$, it is not clear which scheme is better based on current data set.

The re-scale factors of primary CR nitrogen nuclei ($\cnp$) in both schemes are compatible with the corresponding re-scale factors of other primary CR species ($\che$, $\cc$, and $\co$). On the other hand, the re-scale factors of secondary CR nitrogen nuclei ($\cns$) in both schemes are close to 1.0, which are similar to the values of $\cb$ and are the most reasonable values in those $\csec$s. The modulation potential of nitrogen ($\phin$) in both schemes are also compatible with other species.

\begin{table*}[!htbp]
\caption{
The constraints on the parameters of $\cnp$, $\cns$ and $\phin$ based on the best-fit result of Scheme I in Section \ref{sec:fitting_results}. The prior interval, best-fit value, statistic mean, standard deviation and the allowed range at $95\%$ CL are listed. With $\chi^{2}/d.o.f = 25.46/63 $ for best-fit result.}
\begin{center}
\begin{tabular}{lllll}
  \hline\hline
ID  &Prior & Best-fit &Posterior mean and   &Posterior 95\%    \\
    &range&value  &Standard deviation & range  \\
\hline
$\cnp$                            & [0.1, 5.0] & 0.63  &0.64$\pm$0.08   & [0.56, 0.69]\\

$\cns$                            & [0.1, 5.0] & 1.06   &1.06$\pm$0.04   & [1.01, 1.12]\\

$\phin\ (\GV)$                  & [0, 1.5]   & 0.72  &0.72$\pm$0.02   & [0.70, 0.75]\\

\hline\hline
\end{tabular}
\end{center}
\label{tab:nitrogen_scheme_I}
\end{table*}

\begin{table*}[!htbp]
\caption{
The constraints on the parameters of $\cnp$, $\cns$ and $\phin$ based on the best-fit result of Scheme II in Section \ref{sec:fitting_results}. The prior interval, best-fit value, statistic mean, standard deviation and the allowed range at $95\%$ CL are listed. With $\chi^{2}/d.o.f = 23.22/63 $ for best-fit result.}
\begin{center}
\begin{tabular}{lllll}
  \hline\hline
ID  &Prior & Best-fit &Posterior mean and   &Posterior 95\%    \\
    &range&value  &Standard deviation & range  \\
\hline
$\cnp$                            & [0.1, 5.0] & 0.63  &0.05$\pm$0.08   & [0.57, 0.69]\\

$\cns$                            & [0.1, 5.0] & 1.11  &1.11$\pm$0.04   & [1.06, 1.17]\\

$\phin\ (\GV)$                   & [0, 1.5]   & 0.67  &0.68$\pm$0.02   & [0.64, 0.70]\\

\hline\hline
\end{tabular}
\end{center}
\label{tab:nitrogen_scheme_II}
\end{table*}

\section{Discussions and Conclusions}
\label{sec:dis_con}

With the newly released data from AMS-02 \citep{AMS02_proton,AMS02_helium,AMS02_He_C_O,AMS02_pbar_proton,AMS02_Li_Be_B,AMS02_N}, we studied the origin of the hardening in both the primary  (proton, helium, carbon, oxygen, and the primary components of nitrogen) and secondary (anti-proton, lithium, beryllium, boron, and the secondary components of nitrogen) CR nuclei spectra based on two different schemes. Global fitting results have shown that, both of the 2 schemes could have good fitting on current primary and secondary nuclei spectra from AMS-02, and could reproduce the hardening of the spectra obviously.  Moreover, based on current AMS-02 nuclei data, we could not distinguish whether the hardening in these nuclei spectra comes from the sources or propagation. Note that in Figure \ref{fig:secondary_results}, it is obvious that in Scheme II, the predicted spectra of the secondary nuclei are harder than that in Scheme I when $\R \gtrsim 1 \TV$. As a result, more precise secondary nuclei data on high energy/rigidity regions ($> 1 \TV$) is needed to distinguish these two schemes.

\subsection{About $\phi_{i}$s}
In our global fitting, each kind of the species has been employed an independent solar modulation potential $\phi_{i}$ to account for the effects of solar modulation, which is based on the force-field approximation. As a widely used and effective treatment on solar modulation, such configuration could show us its limitations  and the differences of the propagation between different species in the heliosphere.

\begin{figure*}[!htbp]
  \centering
  \includegraphics[width=0.7\textwidth]{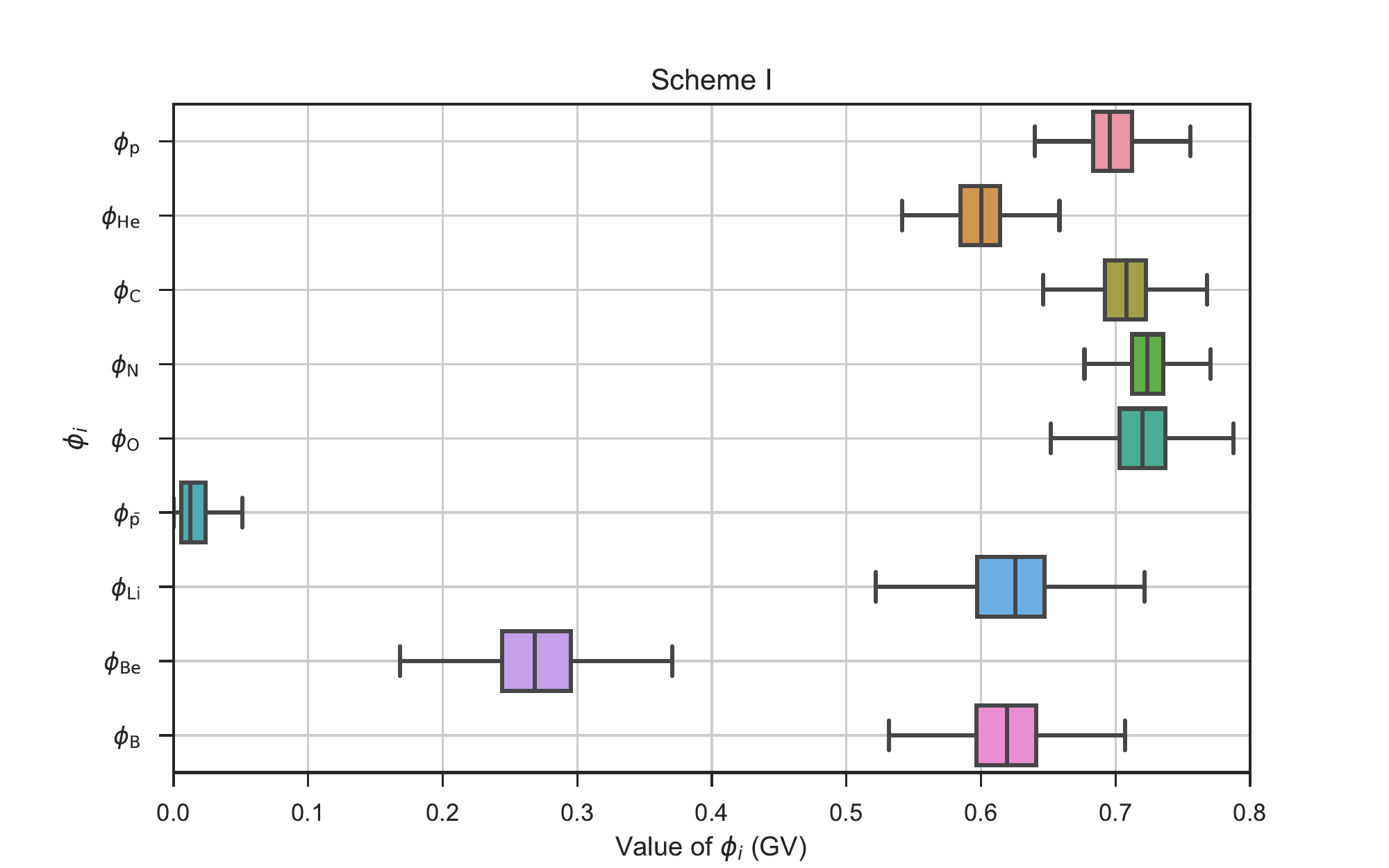}
  \includegraphics[width=0.7\textwidth]{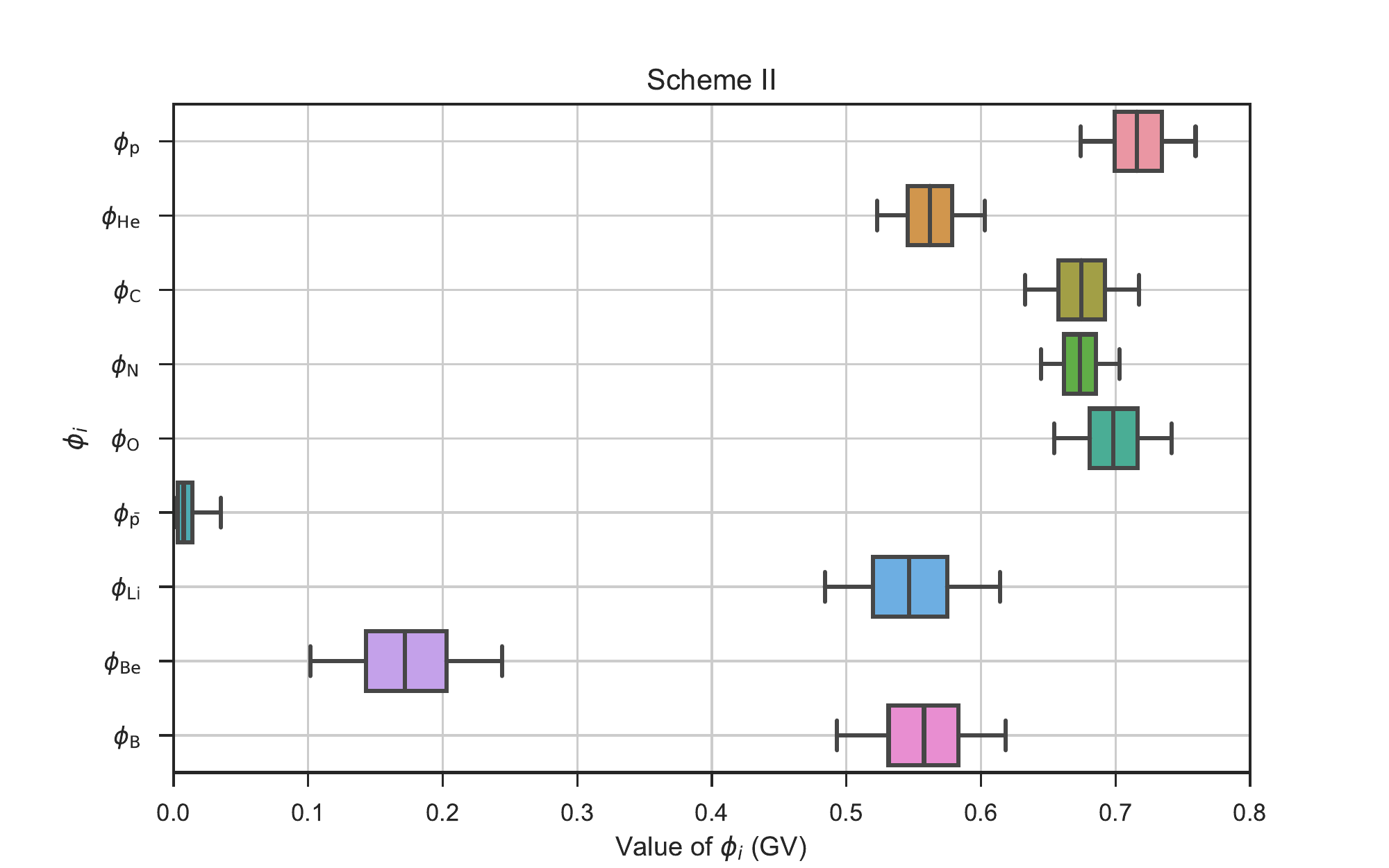}
  \caption{The boxplot for the solar modulation potentials ($\phi_{i}$s) in Scheme I and II.}
\label{fig:phi_box}
\end{figure*}

In order to get a clear representation of the fitting results, we use a boxplot\footnote{A box plot or boxplot is a method for graphically depicting groups of numerical data through their quartiles. In our configurations, the band inside the box shows the median value of the dataset, the box shows the quartiles, and the whiskers extend to show the rest of the distribution which are edged by the 5th percentile and the 95th percentile.} to show  all the $\phi_{i}$ for Scheme I and II in Figure \ref{fig:phi_box}.

In most of the situations, it could give us an acceptable result.\footnote{Except the situation for anti-proton flux, it comes from the charge-sign dependence of the solar modulation, which cannot be handled by force-field approximation.} If we want to study the fine structures in low-energy regions of the spectra ($\lesssim 30 \GV$), we should consider more effects in solving the Parker transport equation which contains diffusion, convection, particle drift and energy loss (see, e.g., \citet{Helmod2017}). On the other hand, the different fitted values of $\phi_{i}$ for different nuclei species indicate some of the species really experience different physical processes in the heliosphere. Especially for beryllium, it needs further researches to reveal the physics behind the value of $\phibe$. \footnote{Here we exclude another special species -- anti-proton, whose particularity would mainly be generated by its negative charge.}

\subsection{About the $c_{i}$s}

\begin{figure*}[!htbp]
  \centering
  \includegraphics[width=0.7\textwidth]{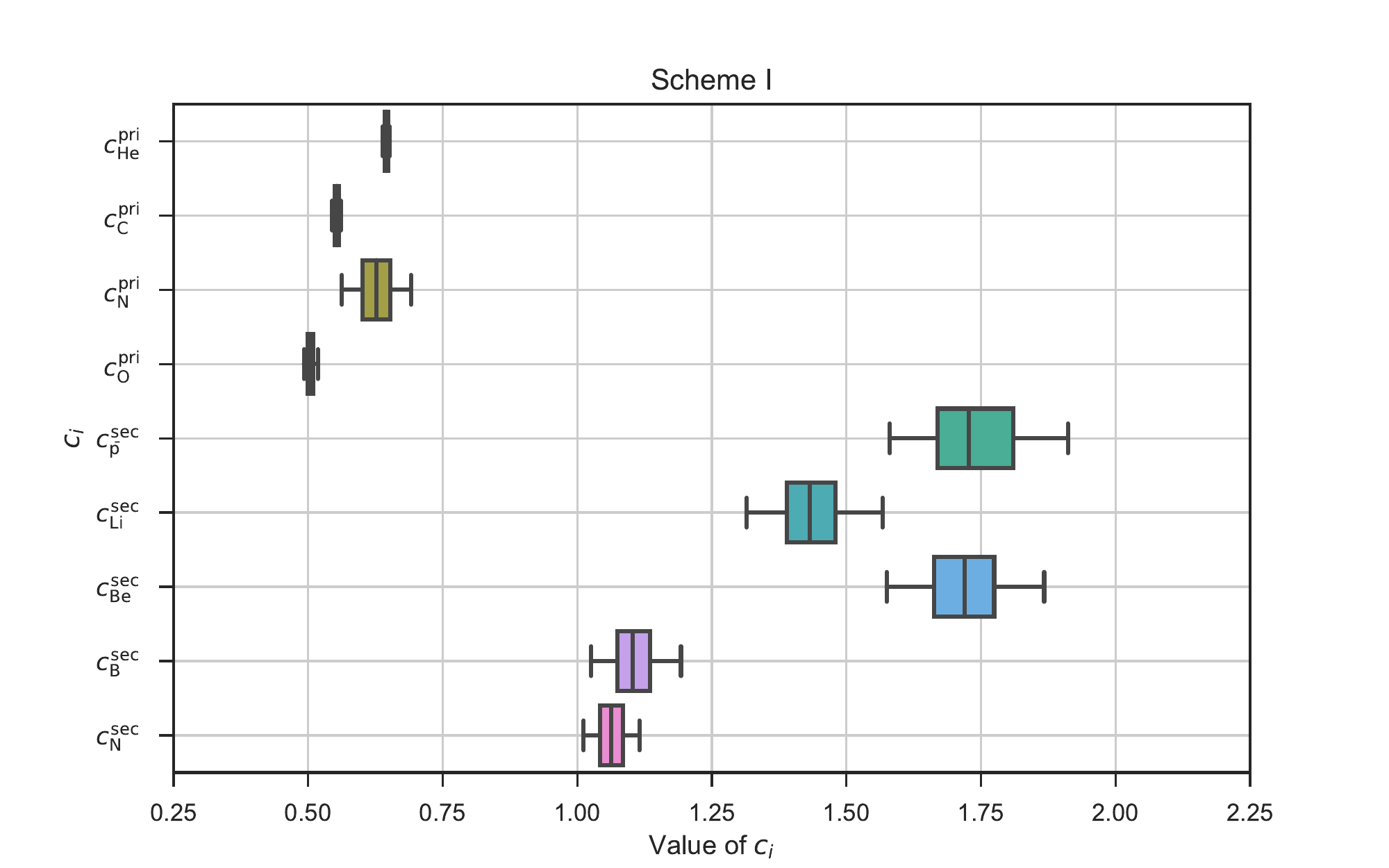}
  \includegraphics[width=0.7\textwidth]{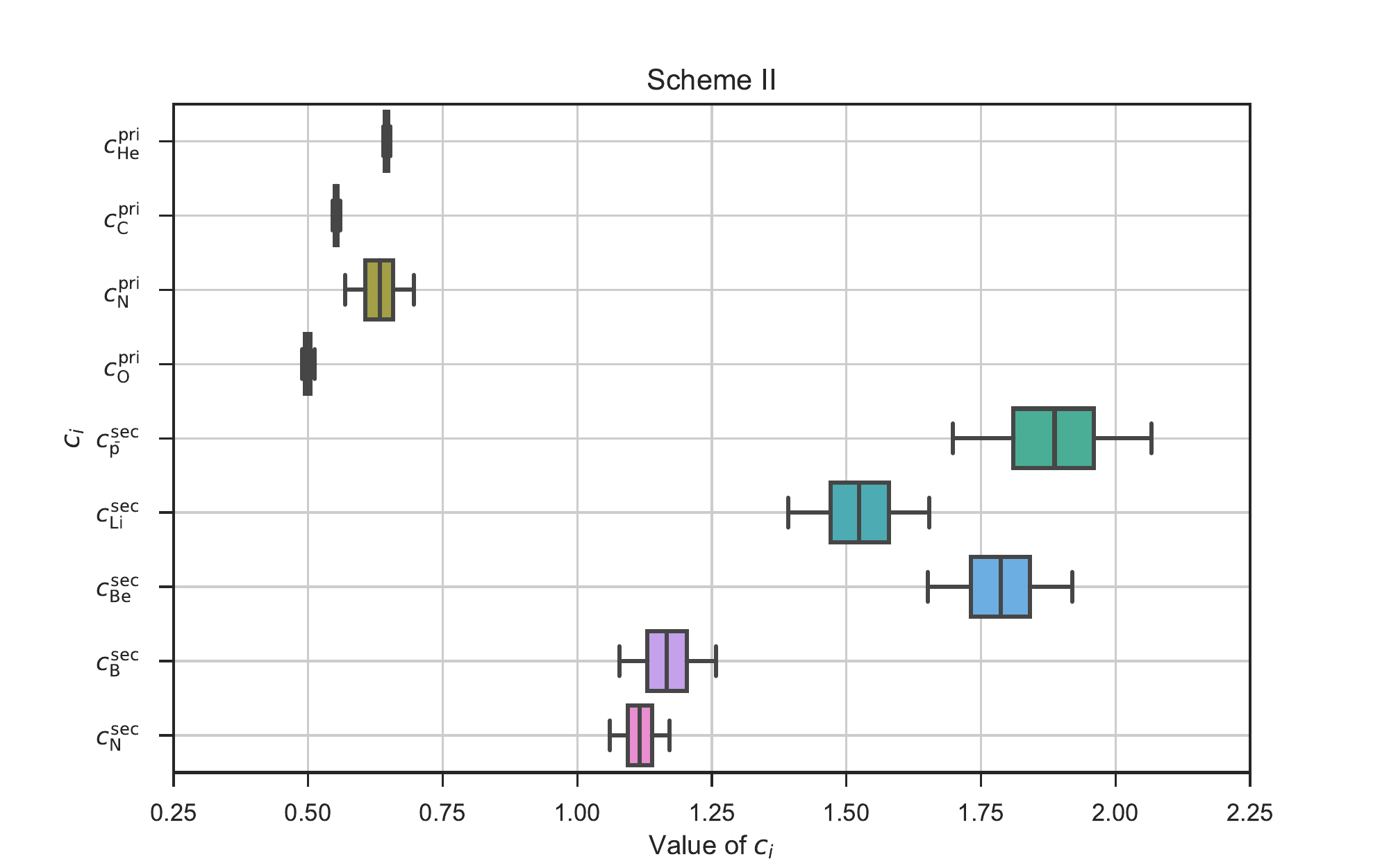}
  \caption{The boxplot for the re-scale factors ($c_{i}$s) in Scheme I and II.}
\label{fig:c_box}
\end{figure*}

Another interesting aspect comes from the fitting values of the re-scale factors.  For convenience, the boxplot of all the $c_{i}$s in Scheme I and II are shown in Figure \ref{fig:c_box}.\footnote{Here, we remove the parameter $N_{\p}$ and show the results of the $c_{i}$s (which are called re-scale factors in this work). } It is clear that, all the primary nuclei re-scale factors have values (which represent their relative element abundances) of $\sim (0.5 -  0.6)$. We know that in {\sc galprop}, the primary source (injection) isotopic abundances are taken first as the solar system abundances, which are iterated to achieve an agreement with the propagated abundances as provided by ACE at $\sim$ 200 MeV/nucleon \citep{Wiedenbeck2001,Wiedenbeck2008} assuming a propagation model. As a result, it is natural that the abundance of the CR species in the solar system (relatively low-rigidity CR particles) is different from that in outer spaces (relatively high-rigidity CR particles). This provides us an effective and independent way to study the isotopic abundance out of the solar system.  Considering the definition of the relative abundance in {\sc galprop}, the re-scale factors of the primary nuclei species which are systematically smaller than 1.0 could be interpreted as: (i) outer spaces have a higher proton abundance than solar system; (ii) outer spaces have a lower abundances of other primary nuclei species (He, C, N, and O) than solar system. All the element abundances we got in this work are listed in Table \ref{tab:abundances}.

\begin{table*}[!htbp]
\caption{Abundances in default {\sc galprop} v56 configuration file and which we got in this work. }
\begin{center}
\begin{tabular}{l|lll}
  \hline\hline
Species  &Default in {\sc galrop}  &Best-fit results in Scheme I  &Best-fit results in Scheme II  \\
\hline
  proton &$10^{6}$ &$10^{6}$  &$10^{6}$ \\
  helium-4  &$7.199 \times 10^{4}$  &$4.629 \times 10^{4}$  &$4.636 \times 10^{4}$ \\
  carbon-12 &$2.819 \times 10^{3}$  &$1.553 \times 10^{3}$  &$1.553 \times 10^{3}$  \\
  nitrogen-14  &$1.828 \times 10^{2}$  &$1.152 \times 10^{2}$  &$1.152 \times 10^{2}$  \\
  oxygen-16  &$3.822 \times 10^{3}$  &$1.926 \times 10^{3}$  &$1.900 \times 10^{3}$  \\

\hline\hline
\end{tabular}
\end{center}
\label{tab:abundances}
\end{table*}

 On the other hand, all the re-scale factors of the secondary nuclei species are larger than 1.0, some of them can reach up to $1.7 \sim 1.8$ ($\cpbar$ and $\cbe$). When we employed these $\csec$s, we expected them to describe the production cross section uncertainties and local inhomogeneity of the CR sources and propagation. Considering the fitting results of these $\csec$s, it seems that the contribution from the production cross section is small, this is because: (i) generally speaking, the production cross sections of these species are energy dependent. But in Figure \ref{fig:secondary_results}, all the CR secondary spectra are well fitted. It is unnatural that all these species have an energy independent correction on their production cross sections; (ii) It is also unnatural that we underestimate all the production cross section of these secondary species simultaneously. Consequently, the fitting results of the $\csec$s (which are systematically larger than 1.0) could be explained as: (i) the ISM density of the outer spaces is larger than that in lcoal environment. This could be regarded as an evidence of the Local Bubble, which the solar system locate in and has a lower ISM density compared with its surroundings  \citep{Lallement2003}; (ii) These secondary nuclei species can be produeced in the CR sources before propagation, which provide an additional flux (see, e.g., \citet{Berezhko2014,Mertsch2014}).

Taking off the systematic deviations, we could find that $\cpbar$ and $\cbe$ still have large deviations compared with other $\csec$s. This might be mainly ascribed to their production cross section uncertainties, which would lead to further studies on these cross sections on colliders and open a new door to study nuclear physics. \footnote{One such pioneering work can be found in \citet{Genolini2018}.}

{\bf Note}: Excluding the fitted values of $\cpbar$ and $\phipbar$ for anti-proton's negative charge, we find that the most special species is beryllium, not only its propagation in the heliosphere, but also its production cross section. This would be related to some interesting problems in stellar physics and cosmology on such special element, which needs further research based on more precise CR data.

\section*{ACKNOWLEDGMENTS}
We would like to thank Qiang Yuan very much for helpful discussions, \citet{Maurin2014} to collect database and associated online tools for charged cosmic-ray measurements, and  \citet{corner} to provide us the tool to visualize multidimensional samples using a scatterplot matrix.
JSN want to appreciate Yi-Hang Nie and Jiu-Qin Liang for their trusts and supports.
This research was supported by the Projects 11475238, 11647601, and 11875062 supported by the 
National Natural Science Foundation of China, and by the Key Research Program of Frontier Science, CAS.
The calculation in this paper are supported by HPC Cluster of SKLTP/ITP-CAS.




\end{document}